\documentclass[twocolumn]{aastex631}

\usepackage{amsmath}
\usepackage{breqn}
\usepackage{xcolor}

\usepackage{float}

\accepted{ApJ}

\shorttitle{DM GHOSts: Rotation and morphology}
\shortauthors{Williams et al.}
\usepackage{xcolor}
\usepackage{mathtools}

\newcommand\lsim{\mathrel{\rlap{\lower4pt\hbox{\hskip1pt$\sim$}}
\raise1pt\hbox{$<$}}}
\newcommand{\lcdm}[0]{\text{$\Lambda$CDM }}
\newcommand{\lcdmc}[0]{\text{$\Lambda$CDM, }}

\graphicspath{{./}{figures/}}

\begin{document}

\title{The Supersonic Project: The eccentricity and rotational support of SIGOs and DM GHOSts }

\correspondingauthor{Claire E. Williams}
\email{clairewilliams@astro.ucla.edu}
\author[0000-0003-2369-2911]{Claire E. Williams}
\affil{Department of Physics and Astronomy, UCLA, Los Angeles, CA 90095}
\affil{Mani L. Bhaumik Institute for Theoretical Physics, Department of Physics and Astronomy, UCLA, Los Angeles, CA 90095, USA\\}
\author[0000-0002-9802-9279]{Smadar Naoz}

\affil{Department of Physics and Astronomy, UCLA, Los Angeles, CA 90095}
\affil{Mani L. Bhaumik Institute for Theoretical Physics, Department of Physics and Astronomy, UCLA, Los Angeles, CA 90095, USA\\}
\author[0000-0002-4227-7919]{William Lake}
\affil{Department of Physics and Astronomy, UCLA, Los Angeles, CA 90095}
\affil{Mani L. Bhaumik Institute for Theoretical Physics, Department of Physics and Astronomy, UCLA, Los Angeles, CA 90095, USA\\}

\author[0000-0003-4962-5768]{Yeou S. Chiou}
\affil{Department of Physics and Astronomy, UCLA, Los Angeles, CA 90095}
\affil{Mani L. Bhaumik Institute for Theoretical Physics, Department of Physics and Astronomy, UCLA, Los Angeles, CA 90095, USA\\}

\author[0000-0001-5817-5944]{Blakesley Burkhart}
\affiliation{Department of Physics and Astronomy, Rutgers, The State University of New Jersey, 136 Frelinghuysen Rd, Piscataway, NJ 08854, USA \\}
\affiliation{Center for Computational Astrophysics, Flatiron Institute, 162 Fifth Avenue, New York, NY 10010, USA \\}

\author[0000-0003-3816-7028]{Federico Marinacci}
\affiliation{Department of Physics \& Astronomy ``Augusto Righi", University of Bologna, via Gobetti 93/2, 40129 Bologna, Italy\\}

\author[0000-0001-8593-7692]{Mark Vogelsberger}
\affil{Department of Physics and Kavli Institute for Astrophysics and Space Research, Massachusetts Institute of Technology, Cambridge, MA 02139, USA\\}

\author[0000-0001-6246-2866]{Gen Chiaki}
\affiliation{Astronomical Institute, Tohoku University, 6-3, Aramaki, Aoba-ku, Sendai, Miyagi 980-8578, Japan}

\author[0000-0002-0984-7713]{Yurina Nakazato}
\affiliation{Department of Physics, The University of Tokyo, 7-3-1 Hongo, Bunkyo, Tokyo 113-0033, Japan}

\author[0000-0001-7925-238X]{Naoki Yoshida}
\affiliation{Department of Physics, The University of Tokyo, 7-3-1 Hongo, Bunkyo, Tokyo 113-0033, Japan}
\affiliation{Kavli Institute for the Physics and Mathematics of the Universe (WPI), UT Institute for Advanced Study, The University of Tokyo, Kashiwa, Chiba 277-8583, Japan}
\affiliation{Research Center for the Early Universe, School of Science, The University of Tokyo, 7-3-1 Hongo, Bunkyo, Tokyo 113-0033, Japan}










\begin{abstract}

A supersonic relative velocity between dark matter (DM) and baryons (the stream velocity) at the time of recombination induces the formation of low mass objects with anomalous properties in the early Universe. 
We widen the scope of the `Supersonic Project' paper series to include objects we term Dark Matter + Gas Halos Offset by Streaming (DM GHOSts)-- diffuse, DM-enriched structures formed because of a physical offset between the centers of mass of DM and baryonic overdensities. 
We present an updated numerical investigation of DM GHOSts and Supersonically Induced Gas Objects (SIGOs), including the effects of molecular cooling, in high resolution hydrodynamic simulations using the {\tt AREPO} code. 
Supplemented by an analytical understanding of their ellipsoidal gravitational potentials, we study the population-level properties of these objects, characterizing their morphology, spin, radial mass, and velocity distributions in comparison to classical structures in non-streaming regions. 
The stream velocity causes deviations from sphericity in both the gas and DM components and lends greater rotational support to the gas. 
Low mass ($\lsim 10^{5.5}$ M$_\odot$) objects in regions of streaming demonstrate core-like rotation and mass profiles. 
Anomalies in the rotation and morphology of DM GHOSts could represent an early Universe analog to observed ultra-faint dwarf galaxies with variations in DM content and unusual rotation curves.

\end{abstract}

\keywords{Galaxy formation --- Dwarf galaxies --- Galaxy rotation curves --- Galaxy morphology}
\section{Introduction} \label{sec:intro}

According to the standard \lcdm model of structure formation, small overdensities seeded by quantum fluctuations in the homogeneous matter fields of the early Universe grew through gravitational collapse into structures.
Prior to Recombination  ($z\sim1100$), overdensities of baryonic matter were prevented from growing by the strong coupling between the baryonic and photonic fields. 
Dark matter (DM) overdensities, however, were free to collapse.  
By the time of Recombination, when baryons decoupled from radiation,  DM overdensities had grown to five orders of magnitude larger than the baryonic overdensities \citep[e.g.,][]{NB05}.
Once decoupled, baryons collapsed into the significantly larger DM potential wells,  resulting in the formation of structures with a central baryon component inside a larger DM halo \citep[e.g.,][]{Wechsler+18}.

This \lcdm  picture of structure formation is very successful on large scales \citep[e.g.,][]{Springel+05,Vogelsberger+14a,Vogelsberger+14b,Vogelsberger+20,Schaye+15}. 
Uncertainties and tensions remain, however, especially on the scales of faint dwarf galaxies  \citep[e.g,][]{BullockBK+17,Simon+19,Perivo+22}. 
From uncertainties such as the core-cusp challenge \citep[e.g.,][]{Flores+94,Moore+94} to serious tensions such as the observed diversity of rotation curves compared to simulations \citep[e.g.,][]{Oman+15,Oman+19}, challenges to \lcdm at low masses include not only tensions with observations \citep[e.g.,][]{Webb+22} but also discrepancies between different state-of-the-art cosmological simulations \citep[see][for a review]{Sales+22}. 
The ultra-faint dwarf regime is thus expected to be one of the most sensitive probes of models and simulations of structure formation that succeed at the scales of Milky Way-like galaxies and larger mass dwarf galaxies. 
A precise description of the morphologies, dynamical histories, and star-formation histories of ultra-faint galaxies under \lcdm (and other models) will be central to resolving these tensions.

In an effort to refine the physical understanding of \lcdmc  \citet{Tes+10a} pointed out that previous work neglected the
highly supersonic relative velocity ($v_{\rm bc}$) between DM and baryonic overdensities stemming from their five orders of magnitude difference in density. 
At Recombination, the root-mean-square (rms) value of the relative velocity ($\sigma_{vbc}$) was 30 km s$^{-1}$, five times the speed of sound of the baryons at the time. 
This velocity has important consequences for structure formation at small scales in the early Universe.
It is coherent over a few Mpc \citep{Tes+10a}, and on those scales it can be modeled as a stream velocity of a single value. 
{Recently, \citet{Uysal+22} provided an observational estimate of the local value of $v_{\rm bc}=1.75^{+0.13}_{-0.28}\sigma_{\rm vbc} $,  suggesting that this effect was present during the formation of the Milky Way.} 

Subsequent works further explored the early-Universe implications of structure formation in the presence of the stream velocity.
For example, the stream velocity was shown to delay the formation of Pop III stars \citep[e.g.,][]{Stacy+10,Greif+11,Schauer+17a} with impacts on reionization and the 21-cm signal {\citep[e.g.,][]{Visbal+12, McOL12,Munoz+19,Cain+20,Park+21,Long+22}}. 
It also suppresses halo abundance and generates ``empty" halos with low gas content \citep[e.g.,][]{Naoz+11a,Asaba+16}, generating large scale inhomogeneities of galaxies \citep[e.g.,][]{Fialkov+11} and affecting the minimum halo mass that holds most of its baryons \citep[e.g.,][]{Naoz+12}. 
Furthermore, in regions with a large relative velocity, gas accretion onto star-forming dwarf halos is affected -- the gas falls downwind of  halos, and has very low densities  \citep[e.g.,][]{OLMc12}.
The stream velocity was shown to be responsible for reducing the number of low mass, luminous satellite galaxies expected in \lcdmc somewhat resolving an existing tension with observation at the time \citep[e.g.,][]{BD}.
Low mass galaxies in the stream velocity also have colder, more compact radial profiles \citep[e.g.,][]{Richardson+13}. 
Beyond galaxies, the stream velocity was suggested to enhance massive black hole formation \citep[e.g.,][]{TanakaLi+13,Tanaka+14,Latif+14,Hirano+17,Schauer+17}.
In addition, the stream velocity produces supersonic turbulence, which can assist with the generation of early magnetic fields in the Universe \citep{Naoz+13}.

Intriguingly, the stream velocity effect is also expected to induce the formation of objects with anomalous properties in patches of the Universe with non-zero values of $v_{bc}$. 
\citet{Naoz+14} showed that the stream velocity introduces a phase shift between DM and baryon overdensities, which translates to a physical separation between the two components.
Two interesting classes of objects arise from this effect that differ from classical \lcdm objects at the same scales. 
First, for objects at a range of low masses ($\lsim {\rm few} \times 10^6$~M$_\odot$), the spatial offset is so large that the baryonic component collapses outside the virial radius of its parent DM halo entirely, potentially surviving as a DM-deficient bound object. 
\citet{Naoz+14} proposed that these Supersonically Induced Gas Objects (SIGOs) may be the progenitors of globular clusters \citep[e.g.,][]{Naoz+14,Popa+15,Chiou+18,Chiou+19,Chiou+21,Lake+21,Nakazato+22,Lake+22}.

Second, for a range of slightly higher mass objects ($\lsim 10^8$~M$_\odot$), the spatial offset is such that the centers of mass of the baryonic component and the parent DM halo are offset, but the majority of the gas remains inside the DM virial radius \citep{Naoz+14}. 
We term these objects Dark Matter + Gas Halos Offset by Streaming (DM GHOSts). 
These structures consist of both a DM and gas component, unlike SIGOs, which are almost entirely gas.
Compared to their classical \lcdm analogues, DM GHOSts are enriched in DM and highly diffuse, because the stream velocity advects a portion of their gas component out of the halo. 
\citet{Naoz+14} suggested that these objects may be the progenitors of ultra faint or dark-satellite galaxies. 

The Supersonic Project was introduced to investigate the supersonic stream velocity-induced objects and their ties to observed structures. 
Previous studies focused on the formation and evolution of SIGOs  \citep[e.g.,][]{Popa+15,Chiou+18,Chiou+19,Chiou+21,Lake+21,Nakazato+22,Lake+22}.
These simulations attempted to confirm the existence of SIGOs and investigate their connection to globular clusters using only adiabatic and sometimes atomic cooling.
All except \citet{Schauer+21,Nakazato+22} and \citet{Lake+22} neglected the effects of molecular hydrogen cooling.
\citet{Popa+15} and \citet{Chiou+19} placed early constraints on the rotational properties of SIGOs, showing that they are highly elongated structures with seemingly greater rotational support than both DM GHOSts and ``classical" analogs--objects of the same mass in regions without streaming. 
\citet{Chiou+18,Chiou+21} and \citet{Lake+22} focused on the potential for SIGOs to be sites of star formation. 
In a semi-analytic study, \citet{Chiou+19} found that SIGOs occupy a similar part of magnitude-radius space today as the population of observed globular clusters \citep[e.g.,][]{McConnachie12}. 
\citet{Lake+21} extrapolated the large-scale variation of SIGO abundances across the sky, predicting anisotropies in the distribution of gas-rich objects at low masses that could be observed by the \textit{James Webb Space Telescope} (JWST) and binary black hole gravitational-wave sources detectable by gravitational-wave detectors. 

Several recent studies indicate that molecular cooling may play an important role in the evolution of SIGOs and other objects in the stream velocity.
\citet{Glover13} and \citet{Schauer+21} indicate that molecular cooling affects the abundance of gas objects in the early Universe,  and \citet{Nakazato+22} found that SIGOs became more filamentary in their molecular cooling simulations.
\citet{Lake+22} studied the collapse of SIGOs in the context of molecular cooling, drawing an analogy to giant molecular clouds, and found that SIGOs should form stars outside DM halos. 
Studies have neither investigated DM GHOSts in detail nor constrained the rotational and morphological properties of the supersonically-induced objects with molecular cooling. 

Here, we present an updated analysis of the morphology, rotation, rotational curves, and mass distribution of both SIGOs and DM-GHOSts using molecular hydrogen cooling numerical simulations supplemented by an analytical perspective.   
We characterize the population-level properties of these elongated objects in the context of ellipsoid potentials, and quantify their total angular momentum and rotational support.
We find that the DM component deviates from a spherical configuration in the presence of the stream velocity.
We also present the first rotation curves for these objects, finding a bifurcation in rotation curve shape according to mass. 
This may serve as an early universe analog to the rotational curve diversity observed in dwarf galaxies \citep[e.g.,][]{Sales+22}. 

The paper is organized as follows: \S~\ref{sec:numerical} describes the numerical simulations used in the study and the classification criteria for SIGOs and DM GHOSts. 
\S~\ref{sec:ellipse} is devoted to the analytical and numerical results of the study. 
In \S~\ref{sec:analytical}, we present the analytical ellipsoid potentials used to understand supersonically-induced objects, and we show the population level morphological properties of SIGOs and DM GHOSts from our simulations in \S~\ref{sec:nummorphology}.
In \S~\ref{sec:spinparam}, we discuss the rotational support and angular momentum of these  objects. 
In \S~\ref{sec:rotation}, we present density profiles and rotation curves of DM GHOSts. 
A summary and discussion of the results is given in \S~\ref{sec:discussion}. The appendices explain the choice of cutoff gas fraction used to define a SIGO (App.~\ref{ap.gasfraction}), a full derivation of the potential and total mass from \S~\ref{sec:analytical} (App.~\ref{ap:potential}), and supplemental morphological data, including comparisons to NFW profiles (App.~\ref{ap:morphology}) . 

In this study we assume a \lcdm cosmology, with $\Omega_{\rm \Lambda} = 0.73$, $\Omega_{\rm m} = 0.27$, $\Omega_{\rm b} = 0.044$, $\sigma_8  = 1.7$, and $h = 0.71$.
{ The value of $\sigma_8  = 1.7$ adopted here is discussed in \S~\ref{sec:objects}.}


\section{Numerical Set Up} \label{sec:numerical}
In a similar manner to previous studies by the Supersonic Project \citep[e.g.,][]{Chiou+18,Chiou+19,Chiou+21}, we perform hydrodynamical simulations using the {\tt AREPO} code \citep[][]{Springel2010a}.
{\tt AREPO} is a moving-mesh code that allows for high resolution studies of structure formation with an accurate picture of the stream velocity up to $z\sim20$.

\subsection{Simulation and Initial Conditions}
We use a modified CMBFAST code \citep[][]{1996ApJ...469..437S}, as presented in \citet{Popa+15}, to include the first-order correction of scale-dependent temperature fluctuations on the initial conditions and their transfer functions, following \citet{NB05}. 
This is necessary as the corrections detemine gas fraction in halos at higher redshift \citep[e.g.,][]{NBM,Naoz+10,Naoz+12}.

\citet{Tes+10a} showed that the supersonic relative velocity is coherent on scales of $\sim$few Mpc, so following \citet{Popa+15}, we choose a box size of 2 comoving Mpc, such that the relative velocity can be modeled as a stream velocity.
Evolution of the stream velocity, a second order correction \citep[][]{Tes+10a}, is also included in the transfer functions. 
The simulations begin at $z=200$, when a $2\sigma$ fluctuation in the stream velocity corresponds to 11.8 km s$^{-1}$.
The stream velocity is thus implemented as a boost of 11.8 km s$^{-1}$ to all baryon particles in the $+x$-direction. 
The box of {2 comoving} Mpc contains 512$^3$ DM particles with a mass resolution of $m_{\rm DM}=1.9\times 10^3\text{ M}_\odot$ and 512$^3$ Voronoi mesh cells representing the gas component, with a mass resolution of $m_{\rm gas}=360\text{ M}_\odot$.
Our results are presented at the end of the simulations, $z=20$.

To investigate the effect of the stream velocity, we perform two runs without the stream velocity (i.e., runs in a region of space with a $0\sigma_{v_{bc}}$ fluctuation in the velocity field) and two runs with a value of $v_{bc}=2\sigma_{v_{bc}}$.
 For each set of two runs (with and without the stream velocity), we include molecular (H$_2$) cooling in one and only adiabatic cooling in the other. The inclusion of molecular cooling is described { in \S~\ref{sec:molecular} below. Our chosen value of $v_{bc}=2\sigma_{v_{bc}}$ is similar to the local observed value $v_{\rm bc}=1.75^{+0.13}_{-0.28}\sigma_{\rm vbc} $ \citep{Uysal+22}.}
\begin{table}
\centering
\begin{tabular}{lll}
\hline
\multicolumn{1}{|l|}{Run} & \multicolumn{1}{l|}{$v_{bc}$}              & \multicolumn{1}{l|}{H$_2$ Cooling}
\\ \hline
\multicolumn{1}{|l|}{$0$v} & \multicolumn{1}{l|}{$0$}                      & \multicolumn{1}{l|}{No}             \\ \hline
\multicolumn{1}{|l|}{$2$v} & \multicolumn{1}{l|}{$2\sigma_{v_{bc}}$} & \multicolumn{1}{l|}{No}             \\ \hline
\multicolumn{1}{|l|}{$0$vH$2$} & \multicolumn{1}{l|}{$0$} & \multicolumn{1}{l|}{Yes} \\ \hline
\multicolumn{1}{|l|}{$2$vH$2$} & \multicolumn{1}{l|}{$2\sigma_{v_{bc}}$} & \multicolumn{1}{l|}{Yes} \\ \hline
\end{tabular}
\caption{Simulation Parameters}
\label{Table:Sims}
\end{table}

\subsection{Molecular Cooling} \label{sec:molecular}
To understand the effect of molecular cooling, we perform two runs for each value of the stream velocity ($0\sigma_{v_{bc}}$ and $2\sigma_{v_{bc}}$), one with adiabatic cooling only and one with molecular cooling included. 
We denote the H$_{2}$ cooling runs with ``H2". 
The 0vH2 and 2vH2 runs were also used in \citet{Lake+22}. 
A summary of the runs in this work is given in Tab.~\ref{Table:Sims}.

As in \citet{Nakazato+22} and \citet[][]{Lake+22}, we explicitly account for nonequilibrium chemical reactions and radiative cooling in the gas, using  GRACKLE, a chemistry and cooling library
\citep[][]{Smith+17,Chiaki+19}.
The 0vH2 and 2vH2 runs include H$_2$ and HD molecular cooling.
The radiative cooling rate of the former includes both rotational and vibrational transitions \citep[][]{Chiaki+19}. 
Chemistry for the following 15 primordial species is included in H2 runs: 
e$^-$, H, H$^+$, He, He$^+$, He$^{++}$, H$^-$, H$_2$, H$_2^+$, D, D$^+$, HD, HeH$^+$, D$^-$, and HD$^+$.
{ We do not include star formation.}

\subsection{Object Classification} \label{sec:objects}
We are interested in  gas-rich structures, including SIGOs, which have somewhat low statistical power in these small box simulations. 
Thus, following \citet{Popa+15,Chiou+18,Chiaki+19,Chiou+21,Lake+22, Lake+21,Nakazato+22} we choose $\sigma_8 = 1.7$, which will increase the statistical power. 
This represents a region of the Universe where structure forms  early, such as in the Virgo cluster \citep[e.g.,][]{NB07}.
{ Because $\sigma_8$  produces a large statistical power, we increase the number of gas objects in the simulation without affecting the cosmology, and these results can then be scaled to other regions accordingly \citep[e.g.,][]{Naoz+12,Park+20}. }

To identify structures, we search for two object classes using a friends-of-friends (FOF) algorithm \citep[see e.g,][]{Popa+15,Chiou+18}. 
\begin{enumerate}
    \item DM-primary/Gas-secondary (DM/G) objects are found using the FOF algorithm on DM particles first. 
    Gas cells in the same object are associated with the DM groups at a secondary stage.
    We require DM/G objects to have at least 300 DM particles, to avoid numerical artifacts. 
    \item Gas-primary (GP) objects are found using the FOF algorithm only on gas cells. 
    This allows us to find objects such as SIGOs in the simulation that have little or no DM component. 
    We require GP objects to have at least 100 gas cells, again in order to avoid non-physical numerical effects. 
\end{enumerate}

\begin{table}
\centering
\begin{tabular}{llll}
\hline
\multicolumn{1}{|l|}{Run} & \multicolumn{1}{l|}{\# GP} & \multicolumn{1}{l|}{\# SIGOs}              & \multicolumn{1}{l|}{\# DM GHOSts}
\\ \hline
\multicolumn{1}{|l|}{$0$v} & \multicolumn{1}{l|}{$2557$}  & \multicolumn{1}{l|}{-}                      & \multicolumn{1}{l|}{-}             \\ \hline
\multicolumn{1}{|l|}{$2$v} & \multicolumn{1}{l|}{$759$} & \multicolumn{1}{l|}{$25$} & \multicolumn{1}{l|}{$734$}             \\ \hline
\multicolumn{1}{|l|}{$0$vH$2$}& \multicolumn{1}{l|}{$5823$} & \multicolumn{1}{l|}{-} & \multicolumn{1}{l|}{-} \\ \hline
\multicolumn{1}{|l|}{$2$vH$2$} & \multicolumn{1}{l|}{$1406$} & \multicolumn{1}{l|}{$69$} & \multicolumn{1}{l|}{$1337$} \\ \hline
\end{tabular}
\caption{Summary of the number of gas primary (GP) objects and subclasses found in the four runs used in this study. Only objects containing over 100 gas particles are included. SIGOs and DM GHOSts do not exist in regions with 0 stream velocity, so they are not tabulated for the 0v and 0vH2 runs, but see App.~\ref{ap.gasfraction} for an investigation of false identification of SIGOs in molecular cooling runs.}
\label{Table:objects}
\end{table}
The choice to cut off DM/G and GP objects at 300 particles and 100 cells respectively gives us a minimum structure mass resolution of $5.7\times10^5$ M$_\odot$ for DM/G and $3.6\times10^4$ M$_\odot$ for GP objects. 

\citet{Popa+15} and \citet{Chiou+18} found that GP objects are often filamentary in nature, and thus a spherical fitting algorithm is {not an optimal choice, as it does not reflect the actual morphology of these structures.}
{We therefore employ the same fitting algorithm of these works, which is based on a triaxial ellipsoid fit.} 
We keep the axis ratio of a triaxial ellipsoid with $N_{0}$ gas particles and maximum radius $R_{\rm max,0}$ around the GP object constant and shrink it in increments of 0.5 percent until the condition $R_{\rm max, n}/R_{\rm max, 0}>N_{\rm n}/N_{0}$ is met, or until $N_{\rm n}/N_{0}<0.8$, where $R_{\rm max, n}$ and $N_{\rm n}$ are the maximum ellipsoid radius and number of gas particles of the $n$th iteration.  
\begin{figure*}

 \center

  \includegraphics[width=\textwidth]{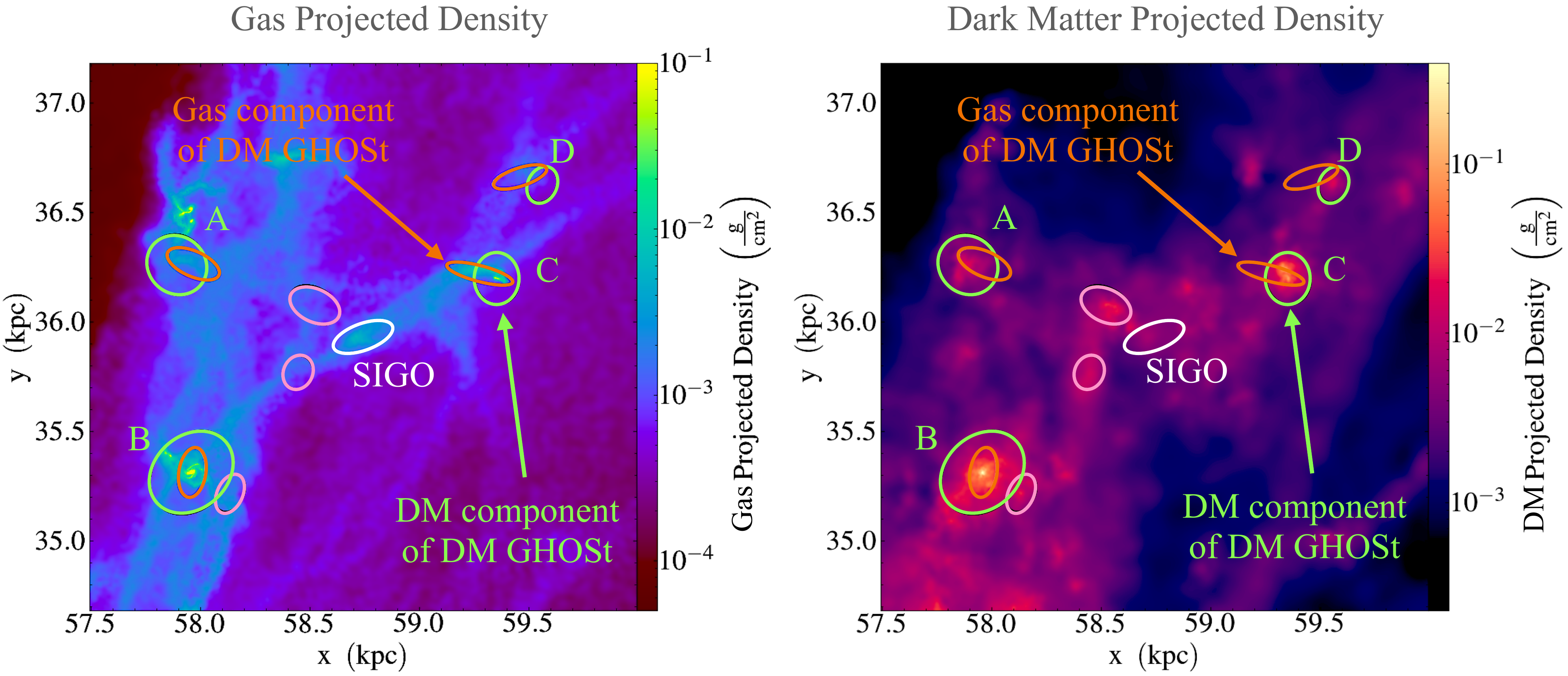}

  \caption{Projected gas (left) and dark matter (DM) (right) density around several DM GHOSts and a SIGO in a region $5$ physical kpc on a side. 
  { The SIGO is bounded by the white ellipse, located in a region relatively devoid of dark matter, and contains no DM component. 
  It is embedded in a stream of gas. 
  The DM GHOSts (A, B, C, and D) each contain a gaseous and a DM component. 
  The gas components of the DM GHOSts are shown in orange, whereas the DM components of the DM GHOSts are shown in green. 
  {The DM components are not entirely spherical.
  The ellipses enclosing the DM components are those found by the method described in \S~\ref{sec:objects} and do not correspond to the DM virial radius of the object, used in the criterion to determine whether an object is a SIGO. }
  The centers of mass of the gas components of A, C, and D are offset from the centers of mass of the DM component, whereas B has had time for the gas component to fall back into the center of the DM potential. 
  Several DM halos with no associated gas components also lie in this region--depicted in pink. 
  One of these may be the ``parent" halo of the SIGO. }
  }
  \label{fig:SIGOdm}

\end{figure*}

The GP FOF algorithm is performed to identify SIGOs, gas-rich objects that form outside the virial radius of the parent DM halos.
However, many of the GP objects are located inside DM halos, being the gas component of the DM/G structures. 
These structures are also of interest to this study.
In order to clarify the difference between structures formed via classical \lcdm and these dark matter and gas structures formed in regions with the stream velocity, we term the DM/G objects in regions of streaming as  Dark Matter + Gas Halos Offset by Streaming (DM GHOSts). 
In previous papers, these were referred to as ``DM/G".
Having formed offset from the center of mass of their parent DM halo, these structures display different morphological and dynamical properties than those {that formed in regions of the Universe with no relative velocity (i.e., a patch with a $0\sigma_{v_bc}$ fluctuation)}, even though many are no longer offset by the redshifts considered here due to dynamical processes {(such as the DM GHOSt labeled ``B" in Fig~\ref{fig:SIGOdm}.)}

We follow the convention in \citet{Nakazato+22}, where SIGOs are defined as GP objects which meet the following two conditions: 
\begin{enumerate}
    \item {They are }located outside the virial radius of their parent DM halo.
    \item {They contain }a gas fraction,
    
\begin{equation}
    f_g=\frac{M_{\rm g}}{M_{\rm DM}+M_{\rm g}} > 0.6 \ ,
    \label{eq.gasfraction}
\end{equation}
where $M_g$ is the total mass of gas in the object and $M_{DM}$ is the total mass of DM in the object. 
\end{enumerate}
Similar criteria were used in \citet{Popa+15,Chiou+18,Chiou+19,Chiou+21,Lake+21}. 
The gas fraction cutoff in those works was chosen rather arbitrarily to be $0.4$. 
This value was implemented because those studies were interested specifically in the gas rich structures in connection with observed DM-deficient objects such as globular clusters.
Our cutoff gas fraction of $0.6$ is higher. 
\citet{Nakazato+22} found that choosing a smaller cutoff gas fraction in runs with molecular cooling leads to the identification of filamentary structure as SIGOs, such that  without the stream velocity many SIGOs are misidentified.
We find similar behavior in our molecular cooling run, and this choice of gas fraction is discussed further in App.~\ref{ap.gasfraction}. 
GP objects in runs with stream velocity that do not meet the SIGO criteria above are classified as the baryonic component of DM GHOSts.   

A DM GHOSt therefore contains two components, a DM component, identified by the DM/G FOF algorithm, and a gaseous component, identified by the GP FOF. 
For DM GHOSts, the GP FOF often identifies the gas structure within the DM-primary object. 
Figure~\ref{fig:SIGOdm} shows the projected density of DM (left) and gas (right) in a region of the simulation box with {four DM GHOSts and a SIGO. }
The SIGO contains only a component identified by the GP FOF, which can be clearly seen in the plot of the gas density. 
The DM GHOSts are found in both particle FOF types and have two overlapping (but offset) components.

In \citet{Chiou+19} and subsequent papers, a spherical overdensity calculation was used to obtain the virial radius of the DM halos. 
However, in this study we sought to explore the morphology of the diffuse DM GHOSts and their dark matter component. 
Thus, we also perform an ellipsoid fit as described above to the DM/G objects to explore whether they deviate from a spherical morphology.
So, as before, a triaxial ellipsoid with fixed axis ratios is fit to the DM/G object, shrinking in $0.5\%$ increments until the axis ratio is less than the ratio of the number of particles in the original object to the shrunken ellipsoid, or $20\%$ of particles are removed.

A table of the GP objects found using the FOF algorithm described here is presented in Tab.~\ref{Table:objects}. 
The probability density distributions in \S~\ref{sec:results} are calculated from this set of objects, with a Gaussian kernel density function using a Scott bandwidth \citep[][]{Scott+2010}. 

\section{Physical properties and Analytical Description}
\label{sec:results}
In combination with an analytical understanding, this section describes the morphological and rotational properties of the population of numerically simulated structures from the four simulation runs described above. 

\subsection{Morphology} \label{sec:ellipse}
Historically, spherical overdensity calculations have been used to understand the gravitational potentials of the Universe's first structures \citep[see][for a review]{BarkanaLoeb+01}. 
However, both the stream velocity and molecular cooling were shown to induce gaseous filaments and elongate structures. 
For this reason,  we introduce the eccentricity as a measure of an object's deviation from an idealized spherical configuration, and present an analytical potential for SIGOs and DM GHOSts in terms of their eccentricity. 
A full derivation of the potential and other relevant equations is given in App.~\ref{ap:potential}.

\subsubsection{Analytical ellipsoid potential of SIGOs and DM GHOSts}
\label{sec:analytical}
\begin{figure}

 \center

  \includegraphics[width=0.4\textwidth]{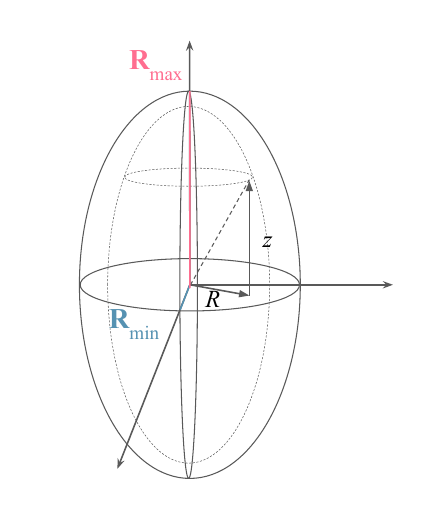}

  \caption{Choice of coordinates adopted in this work.
  The ellipsoids are arranged such that the primary axes of the ellipsoid are aligned along the cartesian coordinate directions, with $R_{\rm max}$, the polar radius of a prolate spheroid, along the $z$-axis. 
  In the prolate approximation, $R_{\rm min}\sim R_{\rm max}$. 
  Cylindrical coordinates are used in \S~\ref{sec:analytical} as the natural choice for prolate ellipsoid potentials. }
  \label{fig:elliosoid}

\end{figure}
In order to analytically explore the role of the eccentricity in the gravitational potential, we approximate SIGOs and DM GHOSts as prolate ellipsoids, with $R_{\rm max}>R_{\rm mid}\sim R_{\rm min}$. We show in Sec.~\ref{sec:nummorphology}  that this approximation is consistent with the structures found in the simulation.   
\begin{figure*}[!t]

 \center

  \includegraphics[width=.95\textwidth]{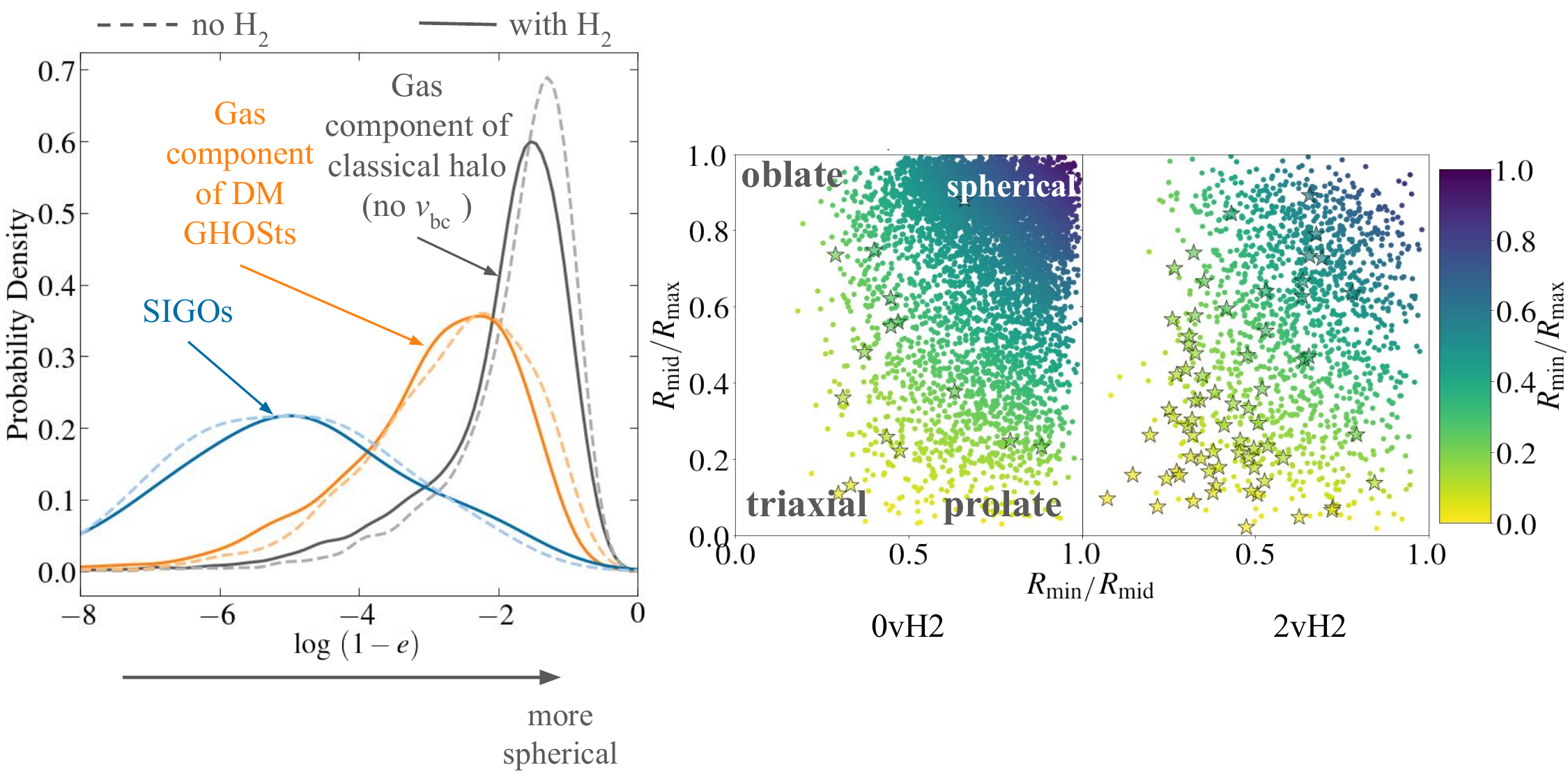}
  
  \caption{ \textit{Left:} Probability density distribution of $\log(1-e)$, where $e$ is the eccentricity (Eq.~(\ref{eq:ecc})), for gas primary (GP) objects.
  Distributions are separated into the object classes listed in Tab.~\ref{Table:objects} and calculated using a Gaussian kernel density. 
 The orange distributions show the gas component of DM GHOSts, the grey distributions show the gas component of classical halos (without $v_{\rm bc}$), and the blue distributions show SIGOs.
 The darker lines denote runs with H$_2$ cooling, while the lighter dashed lines denote no cooling.
 \textit{Right:} Scatter plot of $R_{\rm mid}/R_{\rm max}$ versus $R_{\rm min}/R_{\rm mid}$ for gas-primary (GP) objects in H$_2$ cooling runs.
 The color bar is $R_{\rm min}/R_{\rm max}$.
  The left column has no stream velocity and the right column is from the $v_{bc}=2\sigma_{v_{bc}}$ runs (See Tab.~\ref{Table:objects}). 
  Stars represent SIGOs, as defined in Sec.~\ref{sec:objects}. {In order to make a comparison between the cases with and without the stream velocity, we make an arbitrary cutoff that all three axis ratios must be $\geq 0.9$. For the no stream velocity case, we find 40 objects (0.6\%) that are spherical by this definition in the 0vH2 run and 11 objects (0.4\%) that are spherical in the 0v run. With the stream velocity, we find 0 objects in 2vH2 and 1 object in 2v that meet the criterion. } }
  \label{fig:eccentricity}

\end{figure*}

In cylindrical coordinates ($R,z$), the gravitational potential ($\Phi$) of a prolate ellipsoid can be written as: 
\begin{multline}
 \Phi (\textbf{x}) =\\
    -2^{3/2} \frac{4\pi GR_{\rm max}^4 \rho(R_{\rm max}^2)\cos^{-1}(\sqrt{1-e^2})}{e\sqrt{1+(1-e^2)\left(\frac{R^2}{R_{\rm min}^2}+\frac{z^2}{R_{\rm max}^2}\right)}} \ ,
    \label{eq:cylindricalpotential}
\end{multline}
where $G$ is the gravitational constant,  $R_{\rm max}$ is half the length of the maximum axis of the ellipsoid, $R_{\rm min}$ is half the length of the minimum axis of the ellipsoid (see Fig.~\ref{fig:elliosoid}),  $\rho(R_{\rm max}^2)$ is the density at $R_{\rm max}$, and $e$ is the eccentricity. 
See App.~\ref{ap:potential} for a derivation of Eq.~(\ref{eq:cylindricalpotential}). 
The eccentricity is a measure of the  ellipsoid elongation, defined (following the convention used in \citealp{BT08}) as: 
\begin{equation}
    e\equiv \sqrt{1-\left(\frac{R_{\rm min}}{R_{\rm max}}\right)^2} \ .
    \label{eq:ecc}
\end{equation}
This parameter resembles the 2D ellipse eccentricity, and varies from  0 (spherical) to 1 (radial).
Previous works by the Supersonic Project \citep{Chiou+18} used the prolateness factor ($\xi$) to characterize the shape of GP objects:
\begin{equation}
    \xi = \frac{R_{\rm max}}{R_{\rm min}},
\end{equation}
The relation between the eccentricity and the prolateness factor is:
\begin{equation}
    e = \sqrt{1-\xi^{-2}} \ .
\end{equation}

In deriving the potential in Eq.~(\ref{eq:cylindricalpotential}), we have assumed a prolate spheroidal density profile given by:
\begin{equation}
    \rho(m^2) = \rho_0 \left[1+\left(\frac{m}{R_{\rm max}}\right)^2\right]^{-\frac{3}{2}} \ ,
    \label{eq:density}
\end{equation}
where $m$ is defined in cylindrical coordinates as: 
\begin{equation}
     m^2 \equiv \frac{R^2}{1-e^2}+z^2 \ ,
\end{equation}
\citep{BT08}.
Note that $0\leq m\leq R_{\rm max}$. 
In the above formalism, we scale the density such that $\rho_0 = 2^{3/2} \rho(R_{\rm max})$.

From here, we find the dependence of the total mass on eccentricity.
Once again, a complete derivation can be found in App.~\ref{ap:potential}. 
The total mass of the ellipsoid is found by integrating over a set of similar ellipsoids from the center to the outer edge of the object (i.e., $m=0$ to $m=R_{\rm max}$). 
Thus, for an object with density given by Eq.~(\ref{eq:density}), we find the total mass of the object $M$:
\begin{eqnarray}
    M&=& 4\pi (2^{3/2}\rho_{\rm max})(1-e^2) R_{\rm max}^3 \left(\sinh^{-1}{(1)}-\frac{1}{\sqrt{2}}\right) \nonumber \\
&\approx&     6.19 \rho_{\rm max} (1-e^2)R_{\rm max}^3.
    \label{eq:massecc}
\end{eqnarray}

Below we use the eccentricity parameter to estimate the prolateness of the SIGOs and DM-GHOSts. 
{We also compare the eccentricity inferred by the analytical ellipsoid potential {(Eq.~(\ref{eq:cylindricalpotential}))} for an average object with our simulated objects' eccentricity versus the mass enclosed within the ellipsoid bounding each object, finding agreement between the analytic and numeric results for eccentric objects}.  


\begin{figure*}[!t]

 \center
   \includegraphics[width=.95\textwidth]{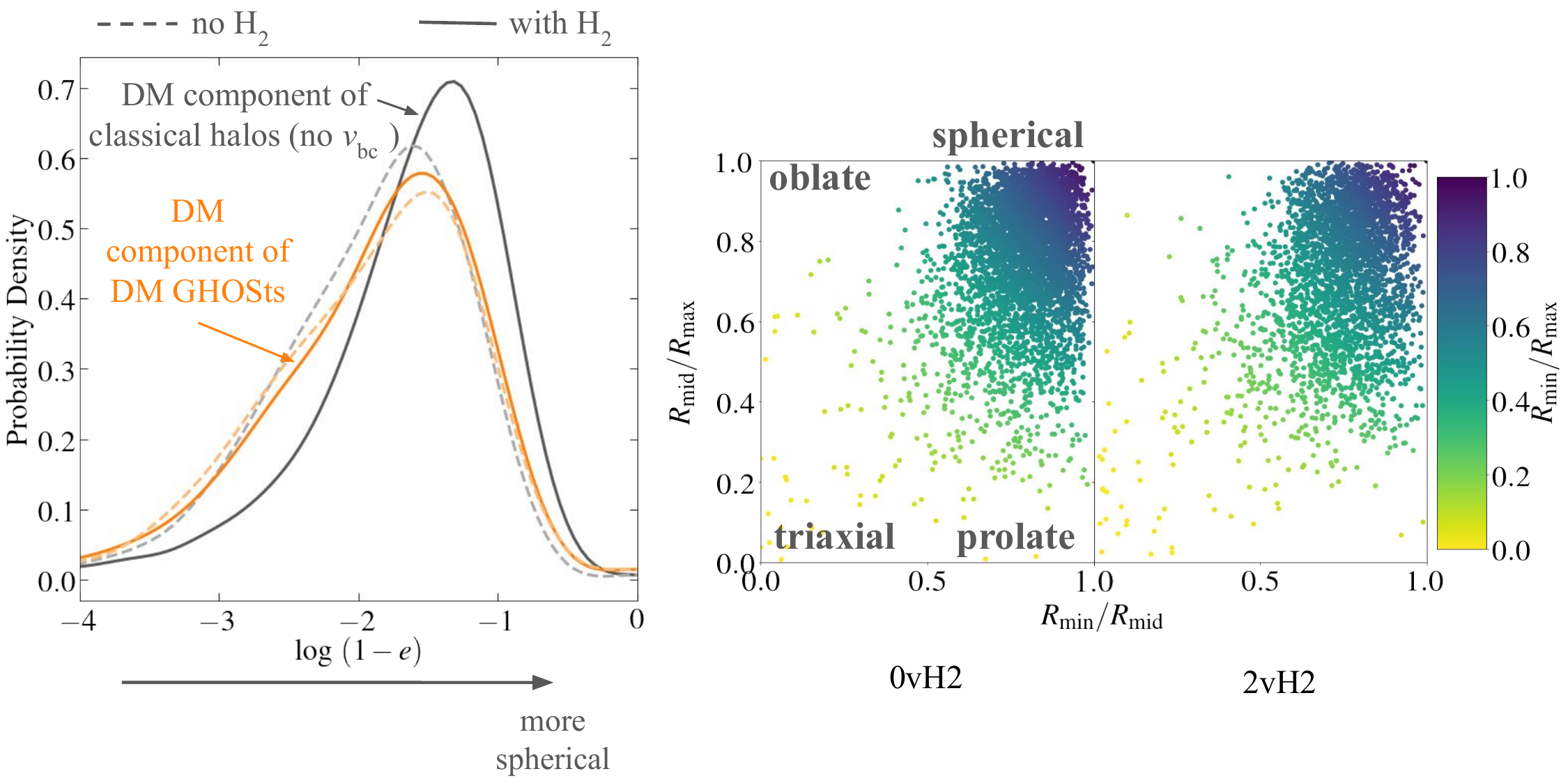}
  \caption{\textit{Left:} Probability density distribution of $\log(1-e)$ of dark matter (DM)-primary objects, where $e$ is the eccentricity (see Eq.~(\ref{eq:ecc})).
 The orange distributions show the DM component of DM GHOSts and the grey distributions show the {DM component of classical halos without $v_{bc}$.}
 SIGOs, which have little to no DM component, are not plotted.
 The darker lines denote runs with H$_2$ cooling, while the lighter dashed lines denote no cooling.
 \textit{Right:} Scatter plot of $R_{\rm mid}/R_{\rm max}$ versus $R_{\rm min}/R_{\rm mid}$ for DM-primary (DM/G) objects.
  The color bar is $R_{\rm min}/R_{\rm max}$.
  The left column has no stream velocity and the right column is from the $v_{bc}=2\sigma_{v_{bc}}$ runs (See Tab.~\ref{Table:objects}). 
  SIGOs are not included because they have no DM component.
  These results imply that the DM component for the majority of objects is non-spherical, and the stream velocity induces further elongation.}
  \label{fig:eccentricityDM}
\end{figure*}
\subsubsection{Morphology of numerically simulated objects}
\label{sec:nummorphology}
\begin{figure}

 \center
  \includegraphics[width=0.48\textwidth]{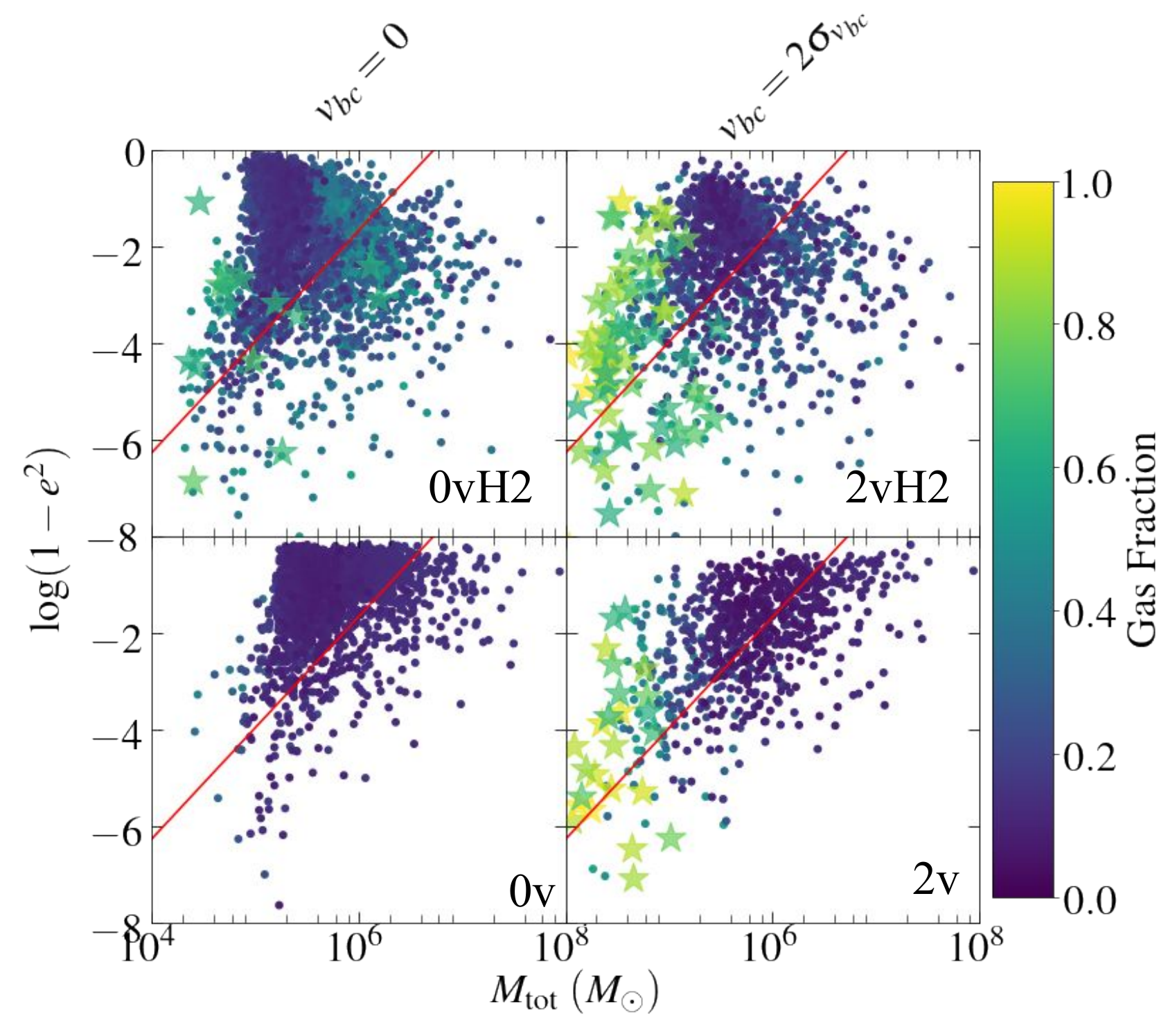}
    \includegraphics[width=0.485\textwidth]{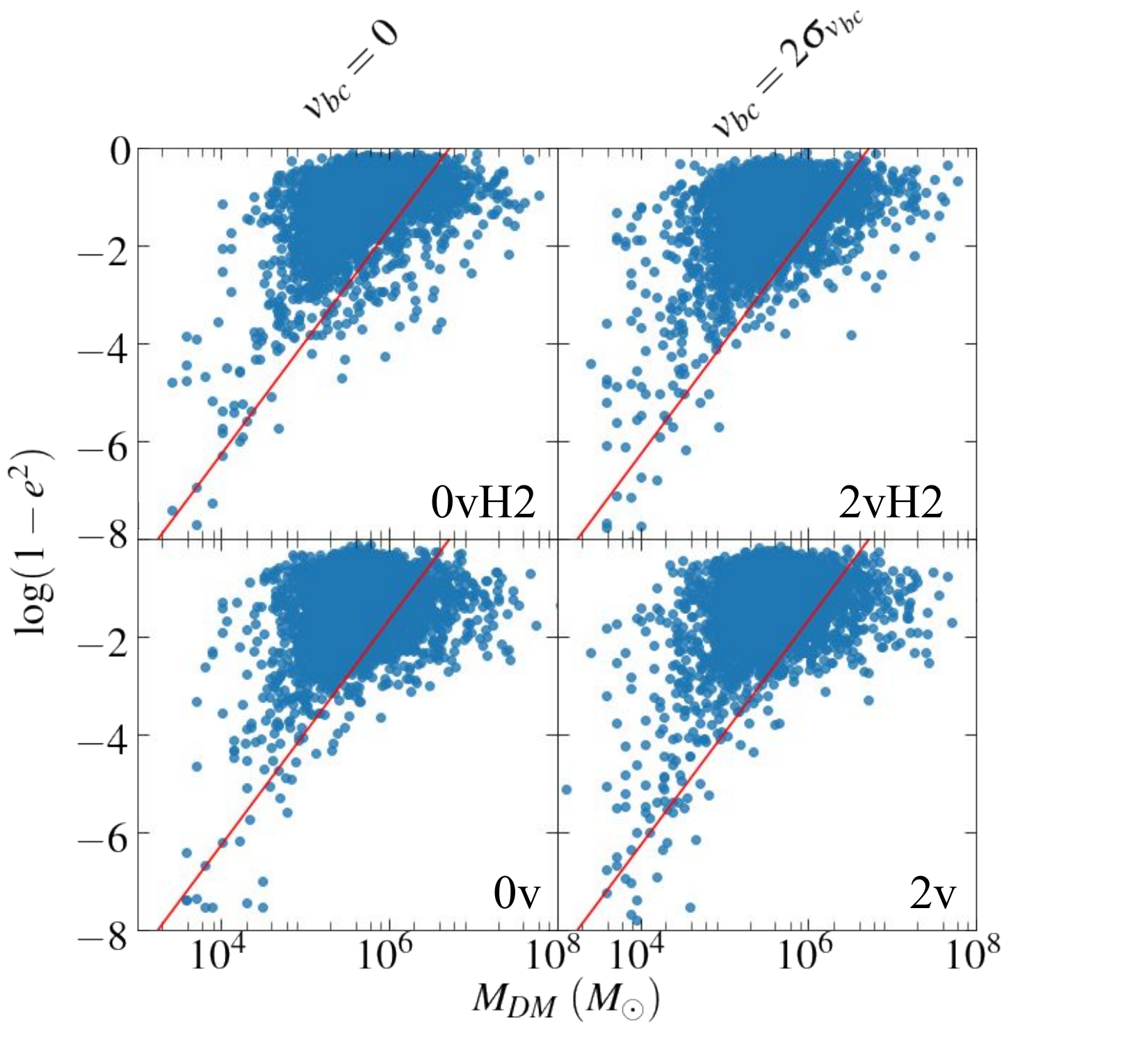}
  \caption{\textbf{Top:} Scatter plot of $\log(1-e^2)$ versus $M_{\rm tot}$ for gas-primary (GP) objects. SIGOs are denoted by stars. 
  \textbf{Bottom:} Scatter plot of $\log(1-e^2)$ versus $M_{\rm DM}$ for DM-primary (DM/G) objects {(we include both DM components of DM GHOSts and other DM halos in the box that have no associated GP component)}
  In both plots, the top two panels show the H$_2$ cooling runs, and the bottom two panels show runs without cooling. 
  The left column has no stream velocity and the right column is from the $v_{bc}=2\sigma_{v_{bc}}$ runs. 
  Stars represent SIGOs, as defined in Sec.~\ref{sec:objects}. 
  The color bar is the gas fraction (Eq.~(\ref{eq.gasfraction})).
  The red overplotted line is the expected relationship from Eq.~(\ref{eq:massecc}) for an example object with the average density and maximum radius of objects in the H$_2$ cooling runs ($\bar{\rho}_{Rmax}=1.8\times 10^8 $M$_\odot$ kpc$^{-3}$, $\bar{R}_{max}=0.134$ kpc).}
  \label{fig:eccmass}

\end{figure}

In the previous section we assumed a prolate relation between ellipsoid axes ( $R_{\rm max}>R_{\rm mid}\sim R_{\rm min}$).
Interestingly, we find that both the gas component and the DM component of structures become prolate in the presence of the stream velocity as depicted in Figure~\ref{fig:eccentricity}. 

The  left panel of Fig.~\ref{fig:eccentricity}  shows probability density distributions for the eccentricity of GP objects in all four runs. 
The gas components of classical halos (i.e., no stream velocity), SIGOs, and DM GHOSts comprise three distinct populations in eccentricity. 
The stream velocity induces elongation of objects, with objects in 0v and 0vH2 runs being the most spherical. 
Among the supersonically-induced objects, SIGOs are more elongated and prolate than DM GHOSts.
The average SIGO eccentricity of $0.977$ corresponds to an object whose $R_{\rm min }$ is only around 20\% of its $R_{\rm max}$, whereas for a classical object with molecular cooling the average ratio is around 60\%
(See Tab. \ref{tab:means} of App. \ref{ap:morphology} for the means of morphological parameters). 
The stream velocity effect dominates runs with and without molecular cooling, but in the no stream velocity case, molecular cooling also slightly elongates the gas component. 

The stream velocity also affects the shape of the DM component of DM GHOSts, resulting in elongated DM structures. 
The  left  panel of Fig.~\ref{fig:eccentricityDM} shows the distribution of eccentricities of the DM-primary objects in the 0v and 0vH2 runs and DM GHOSts with and without cooling. 
Including H$_2$ cooling, the DM component of DM GHOSts tends to be less spherical than the classical DM halos.

The eccentricity is only a measure of the difference between the minimum and maximum radii of the ellipsoid. 
Therefore, to justify the prolate approximation, we show the ratios of all three axes of the gas (Fig.~\ref{fig:eccentricity}, right panel) and DM (Fig.~\ref{fig:eccentricityDM}, right panel) component of the ellipsoids. 
The parameter space is divided into spherical, triaxial, oblate, and prolate objects. 
Even without a stream velocity, there is a range of morphologies among both the DM and gas components of structures. 
The probability density distributions of all the ratios are given in App.~\ref{ap:morphology}.
As seen in the Figures, the majority of the DM components are spherical in nature, and those that deviate from sphericity tend to be prolate. 

For the gas components (top right panel in Figure~\ref{fig:eccentricity}), as expected, the majority of the gas component of classical object is spherical, with preference toward prolate configuration. 
The stream velocity elongates objects into more extreme axis ratios. 
In fact, there are a scarce few truly spherical objects in the runs with stream velocity.
SIGOs, shown as stars in the figure, have not only the most extreme eccentricities overall, but also tend towards the triaxial region of the figure. 

In Fig.~\ref{fig:eccmass}, we plot the eccentricity of the objects' gas component as a function of mass.
We overlay the expected relation from Eq.~(\ref{eq:massecc}) for object of average density and scale (solid line).
Recall that this equation represents the relationship between the mass and the elongation for an ellipsoid potential. 
Thus, for the no stream velocity case, where the majority of the structure is spherical (i.e., $\log(1-e^2)=0$) most of the objects are concentrated at low eccentricity. 
However, more elongated objects in runs with the stream velocity follow the trend outlined by Eq.~(\ref{eq:massecc}). 
In particular, for the no cooling case, the plot shows a more elongated structure for smaller mass systems. 
However, cooling, even in the presence of stream velocity, tends to assist with collapse, thus resulting in a deviation compared to the analytical prediction. 
Note that in general, cooling still tends to create more elongated structures \citep[this was also highlighted in][]{Nakazato+22,Lake+21}.
In the bottom plot of Fig.~\ref{fig:eccmass}, we plot the same relation for the DM/G objects.
Again, those that are more eccentric in nature follow the trend derived from the ellipsoid potential, while many halos, especially classical halos, are concentrated towards circularity.

\begin{figure*}[!t]

 \center

  \includegraphics[width=0.9\textwidth]{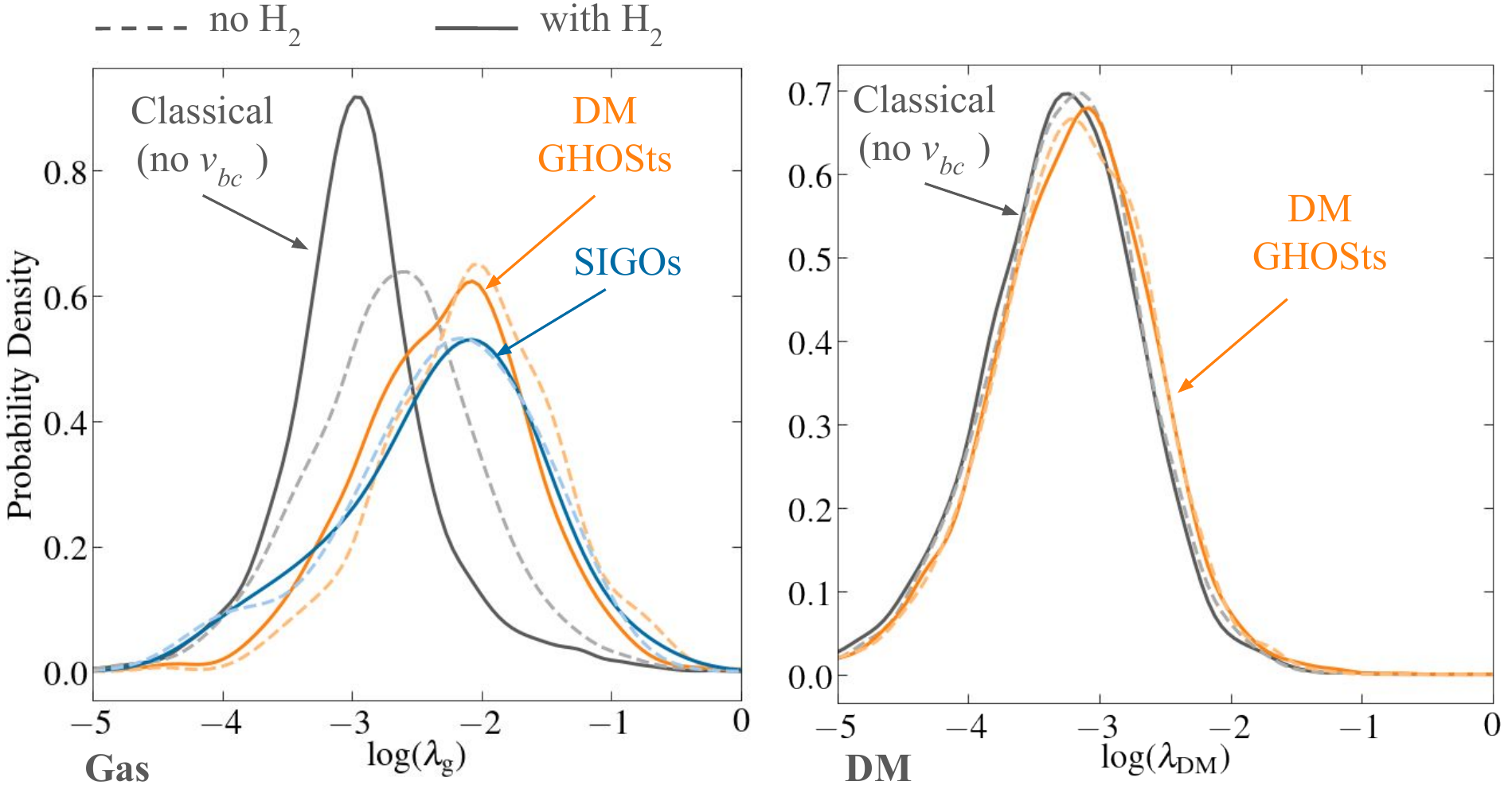}

  \caption{Probability density distribution of $\lambda_{\rm g}$ (left) and $\lambda_{\rm DM}$ (right). 
  These are calculated via Eqns.~(\ref{eq:gasspin}) and~(\ref{eq:spin}), respectively.
 Distributions are separated into the object classes listed in Tab.~\ref{Table:objects} and calculated using a Gaussian kernel density. 
 The orange distributions include the gas component of DM GHOSts, the grey distributions show the gas component of classical halos without $v_{bc}$, and the blue distributions show SIGOs.
 We do not include $\lambda_{\rm DM}$ for SIGOs since they are dominated by gas and the DM does not contribute significantly to the angular momentum of the system. 
 The darker lines denote runs with H$_2$ cooling, while the lighter dashed lines denote no cooling. }
  \label{fig:spins}

\end{figure*}

\subsection{The Spin Parameter} \label{sec:spinparam}

The angular momentum of galaxies has long been understood to be closely tied to their formation and evolution \citep[e.g.,][]{Peebles1969,FallEf80}. 
In particular, the relationship between the angular momentum of the DM halo and the gas seems critical in shaping the final galactic morphology and spin parameter \citep[e.g.,][]{Bullock+01,Maller+02,Danovich+15,RodGom+17,Wechsler+18,Kurapati+21,Yang+21,Rohr+22,Cadiou+22,Ebrahimian+22E,RodGom+22,Hegde+22}.
DM halo spin in simulations follows a lognormal distribution \citep[e.g.,][]{Bullock+01,Zjupa+17}, and the spin of the cold gas of galaxies seems to follow a similar distribution in observations and simulations \citep[e.g.,][]{Danovich+15,Burkert+16}. 
Models that conserve angular momentum suggest that the structure, size, and morphology of galaxies follow the mass and angular momentum of their host halos \citep[e.g.,][]{Somerville+08,Guo+11,Benson+12,Somerville+15}.
Initially, simulations struggled to replicate observed properties of galaxies such as the large spin of the baryonic component compared to the DM and the shape of angular momentum of galaxies, but it was recognized that baryonic processes, including feedback, can explain the evolution of the angular momentum of the baryonic component \citep[e.g.,][]{Maller+02,Teklu+15,Zjupa+17,ElBadry+18,Rohr+22}.
Furthermore, some recent work \citep[e.g.,][]{Sales+12,Danovich+15,Jiang+19} suggests that galaxies' spins are not correlated with the spins of their host halos at all, and observed scaling relations must be explained via other mechanisms. 
Persistent uncertainties in the relationship between angular momentum, morphology, and galaxy structure remain, particularly at low masses \citep[e.g.,][]{Nguyen+22,Ebrahimian+22E}. 
Including $v_{bc}$, which affects both the velocity and the configuration of the baryonic component, has already been shown to affect the spin at low masses \citep[e.g.,][]{Chiou+18}, and thus we continue with an investigation of the angular momentum of our sample of structures. 

To quantify the rotation and angular momentum of objects, we follow the  analytical formulation from  \citet{Chiou+18}.
The total angular momentum, denoted by the spin vector ($\mathbf{J}_{\rm sp}$) of a set of $N$ particles each of mass $m_i$ is
\begin{equation}
    \mathbf{J}_{\rm sp}=\sum_{i=1}^N m_i\mathbf{r}_i\times \mathbf{v}_i,
    \label{eq:spin}
\end{equation}
where $\mathbf{r}_i$ and $\mathbf{v}_i$ are the particles' position and velocity vectors from the center of mass.
For DM primary objects, we estimate the angular momentum of the entire halo using the spin parameter \citep[e.g.,][]{Peebles1969} as defined in \citet{Bullock}:
\begin{equation}
    \lambda_{\rm DM} = \frac{J_{\rm sp}}{\sqrt{2} M_{200}v_{200}R_{200}} \ .
    \label{eq:spinparameter}
\end{equation}
Here $M_{200}$ is the virial mass of the object, $v_{200}= \sqrt{GM_{200}/R}$ \citep[][]{BarkanaLoeb+01}, and $J_{\rm sp}=|\mathbf{J_{\rm sp}}|$.
\citet{Chiou+18} showed that the DM/G spin follows a lognormal distribution consistent with \citet{Bullock+01}.

Following \citet{Chiou+18}, in order to account for the more ellipsoidal nature of the gas component, we calculate the spin parameter for gas primary objects using: 
\begin{equation}
    \lambda_{\rm g} =\frac{J_{\rm g }}{6\sqrt{2}M_{\rm g}v_{\rm GP}R_{\rm max}},
    \label{eq:gasspin}
\end{equation}
where $M_{\rm g}$ is the total gas mass, $v_{\rm GP}$ is the circular velocity of the gas primary object at a distance $R_{\rm max}$, and $J_{\rm g}$ is calculated from Eq.~(\ref{eq:spin}) for gas particles only.

Figure~\ref{fig:spins} shows the probability density distributions of the spin parameter of gas primary objects and DM primary objects. 
The stream velocity induces higher total spin for the gas component of all runs. 
However, for classical gas objects, molecular cooling serves to lower the total angular momentum by condensing gas inward, thus allowing for a smaller spin parameter. 
For classical objects, the DM components have a larger spin parameter magnitude overall: because DM constitutes most of the mass, it contributes most of the  total angular momentum. 
On the other hand,  the stream velocity boosts the gas spin for both SIGOs and the gas component of DM GHOSts, thus increasing the total angular momentum of the system despite H$_2$ cooling. 
SIGOs and DM GHOSts thus have gas spin parameters an order of magnitude larger than the no stream velocity gas (see Tab.~\ref{tab:means}).

\begin{figure*}[!t]

 \center

  \includegraphics[width=\textwidth]{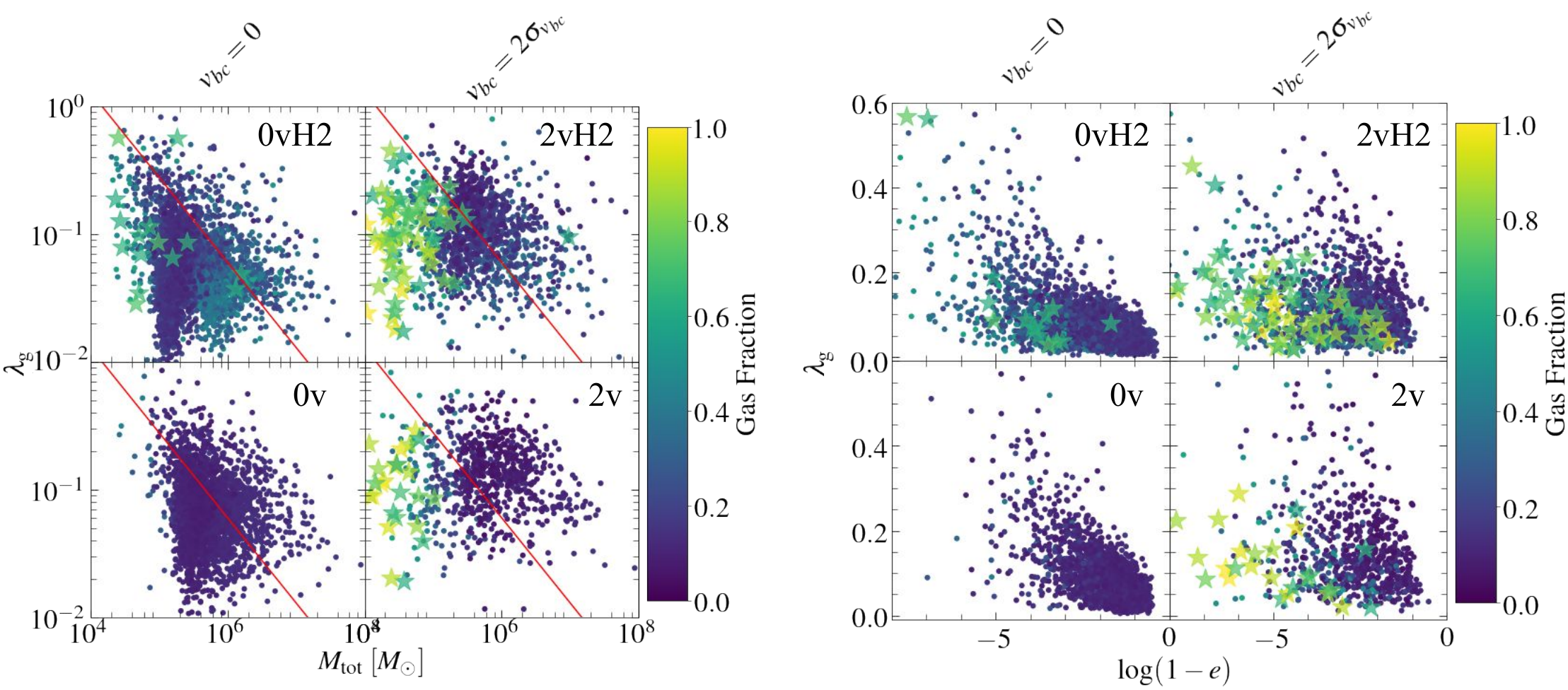}

  \caption{Scatter plot $\lambda_{\rm g}$ (left) versus the dark matter mass ($M_{\rm tot}$) (left) and $\lambda_{\rm g}$ versus $\log(1-e)$ (right)  of GP objects.
  The top row includes H$_2$ cooling. 
  The left column denotes the runs without the inclusion of the stream velocity, and the right column contains runs with the stream velocity.
  A line corresponding to $\lambda_g \sim M^{-2/3}$ is overplotted in red on the right hand side. 
  This is the expected relation from Eq.~(\ref{eq:gasspin}).
  The vertical cut-off at low masses is due to our resolution limit of $300$ particles minimum (corresponding to a mass of $5.7\times 10^5$~M$_\odot$), while at high mass we are limited by Poisson fluctuations of small number statistics at this high redshift and small box size.}
  \label{fig:spinscombined}

\end{figure*}

In Fig.~\ref{fig:spinscombined}, we plot the spin parameter against the total mass (left) and eccentricity (right) of objects.
More eccentric objects tend to have higher spins in all runs. 
The vector sum in the definition of $\mathbf{J}_{\rm sp}$ (Eq.~(\ref{eq:spin})) means that this parameter encodes not only the magnitude of the total angular momentum but also the alignment of particles' rotation.
Thus, the trend on the right of Fig.~\ref{fig:spinscombined} is consistent with spherical configurations corresponding to an isotropic distribution of the particles' orbits. Further, it is consistent with prolate systems having a preferred directionality to the angular momentum or ordered distribution of particle orbits. 

Furthermore, the spin of the objects in the no stream velocity case roughly follows a $\lambda\sim M^{-2/3}$ slope, see bottom left panel in Fig.~\ref{fig:spinscombined}.
This relation is expected from Eq.~(\ref{eq:spinparameter}) for mostly circular orbits. 
However, the trend dissipates in the presence of the stream velocity and cooling, where the objects deviate from spherical symmetry and the combined effects introduce a preferred directionality for the angular momentum, almost regardless of the mass.
We attribute this to the turbulent and filamentary nature of these structures in the presence of the stream velocity and molecular cooling \citep[e.g.,][]{Nakazato+22,Lake+22}.  
Note that the cut-off in the low-mass regime is due to our resolution limit of $300$ particles minimum (corresponding to a mass of $5.7\times 10^5$~M$_\odot$), while at the high mass regime we are limited by Poisson fluctuations of small number statistics at this high redshift and small box size.

The question of whether these larger spin parameters imply greater overall angular momentum or more ordered rotation leads us to an investigation the connection between the morphology of the objects and their rotational support. 
{An investigation of the ellipsoids' alignment with respect to the stream velocity direction revealed that SIGOs are not always aligned with $R_{max}$ in the direction of the stream velocity. 
More frequently, they are embedded in a stream of gas that is infalling towards a larger DM halo (see Fig.~\ref{fig:SIGOdm}, for example), and their longest axis aligns with this stream of gas. 
Thus, to check how ordered these objects rotation is, we test whether the rotation axis is aligned with any of the three primary ellipsoid axes.}

\begin{figure}

 \center

  \includegraphics[width=0.3\textwidth]{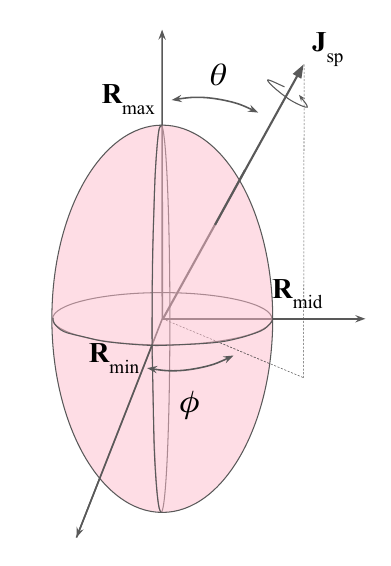}

  \caption{
  Ellipsoid in spherical coordinates for
  an arbitrary 
  spin vector direction.
  Coordinates are chosen such that the primary axes of the ellipsoid are aligned along the cartesian coordinate directions, with $R_{\rm max}$, the polar radius of the prolate spheroid, along the $z$-axis. 
  The spin vector (Eq.~(\ref{eq:spin})) can be aligned in any direction with respect to the ellipsoid axes, and its relative alignment with respect to these axes is described by the usual spherical angular coordinates $\theta$ and $\phi$.  }
  \label{fig:elliosoid2}

\end{figure}
To describe the directionality of the angular momentum we utilize spherical coordinate notation.
With the maximum radius aligned with the $z$-axis and the minimum radius aligned with the $x$-axis, we calculate the spherical $\theta$ and $\phi$ components of the spin vector of both the DM and gas of objects. 
Fig.~\ref{fig:elliosoid2} shows an illustration of the orientation of the ellipsoid with respect to the spin vector. 

\begin{figure*}

 \center

  \includegraphics[width=0.85\textwidth]{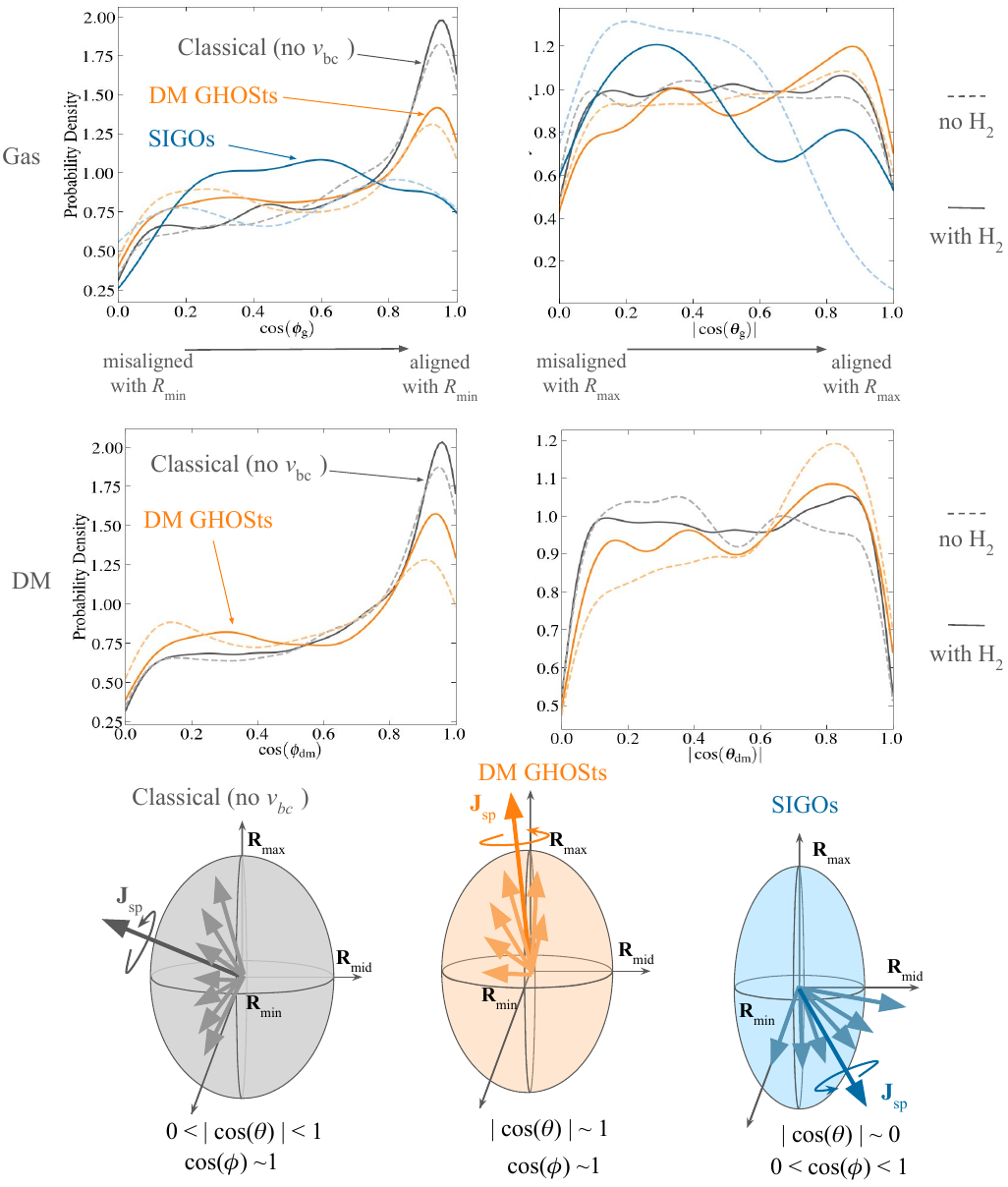}

  \caption{\textbf{Top panels:} \textit{Left:} Probability density distribution of $\cos{(\phi_{\rm g})}$ the angle between $R_{\rm min}$ and the spin vector of the gas component of  GP objects (See Fig.~\ref{fig:elliosoid}). \textit{Right:} Probability density distribution of  $|\cos{(\theta_g)}|$, the angle between $R_{\rm max}$ and the spin vector. 
The darker lines denote runs with H$_2$ cooling, while the lighter dashed lines denote no cooling.
 Distributions are separated into the object classes listed in Tab.~\ref{Table:objects} and calculated using a Gaussian kernel density. 
 The orange distributions include the gas component of DM GHOSts, the grey distributions show the gas component of classical halos without $v_{bc}$, and the blue distributions show SIGOs. 
 \textbf{Middle panels:} \textit{Left} Probability density distribution of $\cos{(\phi_{\rm dm})}$ the angle between $R_{\rm min}$ and the spin vector of the DM component of objects (See Fig.~\ref{fig:elliosoid}). \textit{Right:} Probability density distribution of  $|\cos{(\theta_{\rm dm})}|$, the angle between $R_{\rm max}$ and the spin vector. 
The darker lines denote runs with H$_2$ cooling, while the lighter dashed lines denote no cooling.
 Distributions are separated into the object classes listed in Tab.~\ref{Table:objects} and calculated using a Gaussian kernel density. 
 The orange distributions include the gas component of DM GHOSts, the grey distributions show the gas component of classical halos without $v_{bc}$. 
 SIGOs, which have little to no DM component, are not shown.
 \textbf{Bottom panels:} Cartoon depiction of the orientation of the spin vector with respect to the various object classes, classical halos (left), DM GHOSts (middle) and SIGOs (right). 
 The large, dark arrow indicates the most common orientation, while the other faint arrows indicate a spread of other common alignments for each object type according to the distributions in the top and middle panels.}
  \label{fig:spinsradii}

\end{figure*}
Figure~\ref{fig:spinsradii} shows distributions of the misalignment between the primary ellipsoid axes and the spin vector for both the gas component (top row) and the DM component (middle row). 
As depicted, the classical halos (both for DM and gas components) are preferentially spinning in alignment with their minimum axis, and do not show a preference with respect to their maximum axis. 
These classical halos were the most oblate group overall, thus the lack of preference for alignment with the maximum axis could be due to the fact that for an oblate spheroid, $R_{\rm max}\sim R_{\rm mid}$. 
In other words, they are consistent with puffy disks. 
The bottom row of Fig.~\ref{fig:spinsradii} shows a cartoon depiction of the range of preferred rotation of classical objects here. 

DM GHOSts, to a lesser degree, are rotating in alignment with their minimum axis, but they also show a preference towards the maximum axis. 
This ``spinning top'' type of behaviour seems to be consistent when considering their  formation \citep[see Fig. 3 in][]{Lake+22}. 
The gas component of DM GHOSts (similar to classical objects) is accreted in a stream onto the halo, but the stream velocity induces a velocity gradient in a preferential direction in that region perpendicular to the infall stream.
This results in spinning-top rotator behavior, depicted also in a cartoon at the bottom of Fig.~\ref{fig:spinsradii}.

SIGOs, however, exhibit a weak bimodal distribution of alignment with $R_{\rm max}$.
The majority are preferentially {\it misaligned} with the maximum axis, while some demonstrate alignment as in the case of the DM GHOSts described above. 
Considering an idealized growth scenario, SIGOs are embedded in the gas stream, which is normally in the process of accreting onto a DM halo. 
This configuration often yields an $R_{\rm max}$ in alignment with the accretion stream, (as is the case in the example in Figure \ref{fig:SIGOdm}).
{As described above with DM GHOSts, the stream velocity induces a velocity gradient (in our case towards the $x-$direction) which may be perpendicular to the infall stream moving in the $y-$ or $z-$directions.  \citep[For example, see Fig. 3 in][where a SIGO is embedded in a stream of gas infalling towards a larger halo. All the gas in the region is moving towards this stream, however greater velocities are found on the $+x$ side, a gradient induced by the original stream velocity in the $+x$-direction.]{Lake+22}}
This perpendicular accretion mode causes objects with alignment between $\mathbf{J}_{\rm sp}$ and $\mathbf{R}_{\rm max}$.
However, this picture is idealized, and in practice the SIGOs represent a density perturbation within the stream that results in gas accretion that is not necessarily aligned with the object's $R_{\rm max}$. 
Those SIGOs that are preferentially misaligned respect to $R_{\rm max}$  show a variety of alignments with respect to the $R_{\rm min}$ .  
This is potentially due to the fact that (as opposed to the oblate case described above) for prolate spheroids, the system's symmetry is that $R_{\rm min}\sim R_{\rm mid}$. 

\begin{figure}

 \center

  \includegraphics[width=0.45\textwidth]{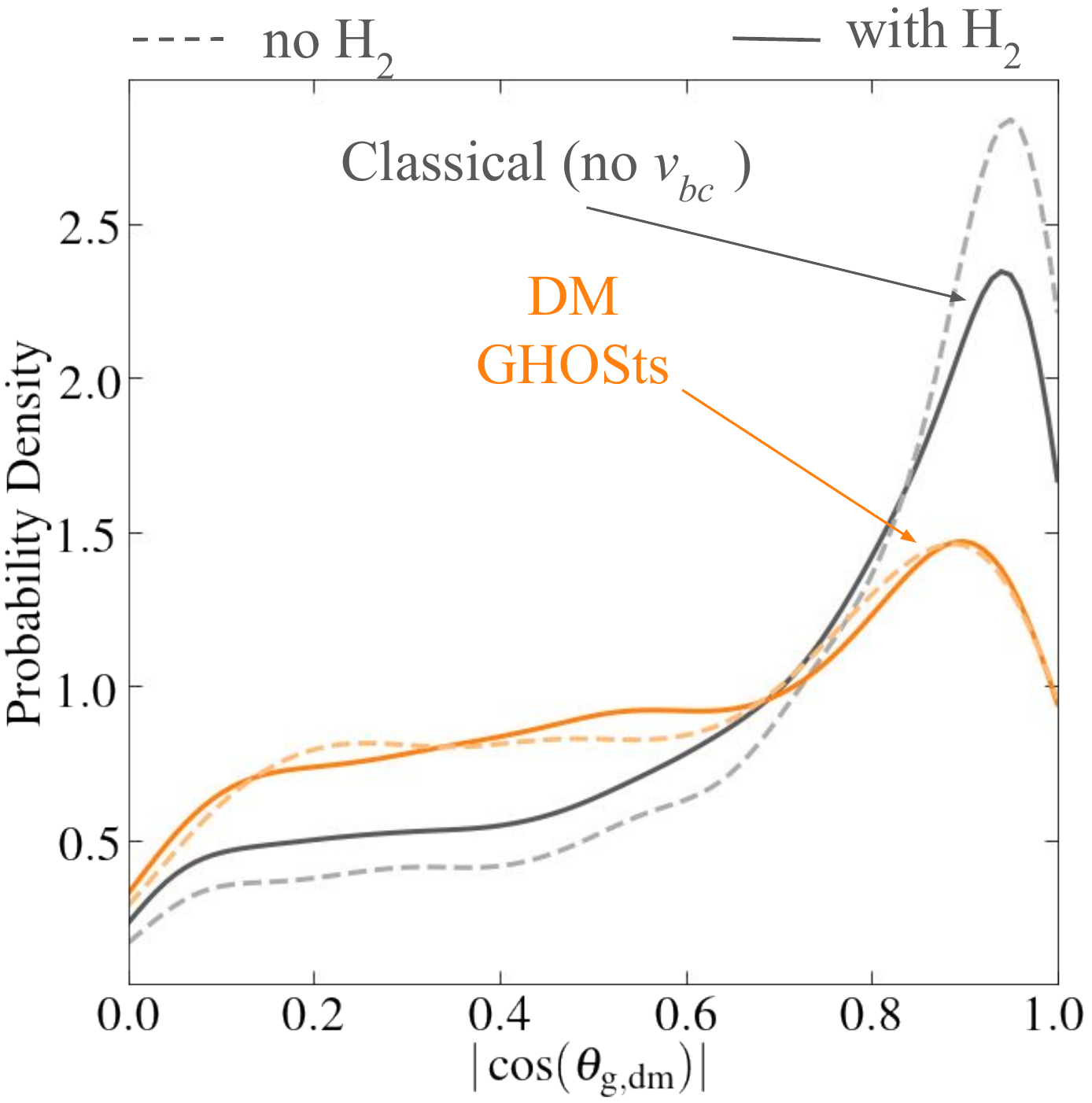}

  \caption{Probability density distribution of the cosine of the misalignment between the spin parameter of the gas and dark matter components of individual GP objects.
  (See Eq.~(\ref{eq:thetamisalign}))
  The grey distributions are classical halos, the orange distributions are DM GHOSts, and the blue dashed distributions show SIGOs.
 The darker lines denote no cooling, while the lighter dashed lines show the inclusion of molecular cooling.
 }
  \label{fig:misalignment}

\end{figure}

The similarity between the  DM and gas components in Fig.~\ref{fig:spinsradii} is further investigated below. Specifically, we calculate the misalignment angle between the angular momentum of the DM component and the gas component: 
\begin{equation}
    \cos{(\theta_{\rm g,DM})}=\frac{\mathbf{J}_{\rm DM}\cdot\mathbf{J}_{\rm g}}{|J_{\rm DM}||J_{\rm g}|} \ .
    \label{eq:thetamisalign}
\end{equation}
Note that for SIGOs the DM component is negligible, thus we only consider the classical objects and DM GHOSts in this analysis. 
Figure~\ref{fig:misalignment} shows the probability distributions of $\cos{(\theta_{\rm g,DM})}$. Consistent with previous studies \citep[e.g.,][]{Chiou+18}, the classical halos are have a strong alignment between the gas and DM components. 
On the other hand, the alignment between gas and DM spin is weaker for the DM GHOSts, with a long tail of nearly isotropic configurations.  
This result is consistent between molecular and atomic cooling. 
In future work it may also be relevant to examine the effects of feedback on this distribution. 
This may especially be relevant for star-forming SIGOs. 

\subsection{Rotation Curves and Mass Distribution of DM GHOSts}
\label{sec:rotation}

Because the stream velocity affects the angular momentum and morphological configuration of structures, we expect a possible effect on the density distribution and rotational curves. 
In particular, since rotation curves contain signatures of the DM component, we expect that both SIGOs and DM GHOSts will deviate from the classical profiles. 

\begin{figure}

 \center

  \includegraphics[width=0.48\textwidth]{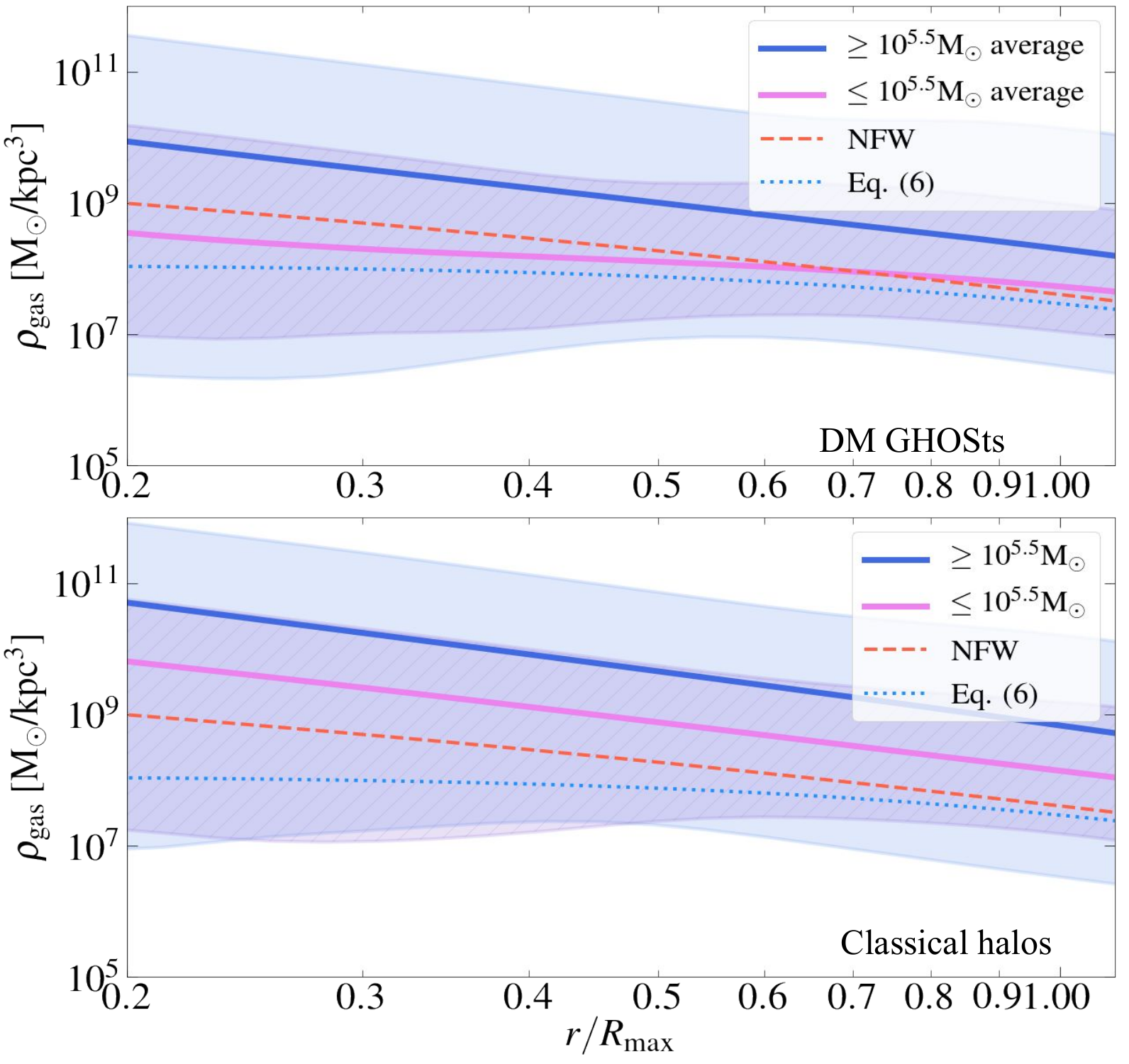}

  \caption{Density of gas in DM GHOSts as a function of radius, normalized to the $R_{\rm max}$ of the ellipsoid, for the molecular cooling run with $v_{bc}=2\sigma_{v_{bc}}$ (top panel) compared to classical halos with $v_{bc}=0\sigma_{v_{bc}}$ and molecular cooling (bottom panel). 
  The density is calculated in 50 ellipsoidal shells moving out from the center of the object.
  We split objects by mass--the average rotation curve for objects above $10^{5.5}$ M$_\odot$ is shown in solid blue, while those below this cutoff are plotted in solid pink. 
  Low mass objects display a deviation from the classical rotation curves, with less of a cusp.
  The blue shaded region shows $1\sigma$ away from the curve for objects above $10^{5.5}$ M$_\odot$, while the purple hatched region shows $1\sigma$ away from the curve for objects below $10^{5.5}$ M$_\odot$. 
  An NFW profile for a $10^5$ M$_\odot$ halo is shown for comparison as the dashed line. 
  Eq.~(\ref{eq:density}) is plotted for a gas object with average density in solid blue. }
  \label{fig:densitymol}

\end{figure}

In particular, in this section we focus our analysis to a comparison between runs with and without stream velocity (0vH2 and 2vH2). 
Figure~\ref{fig:densitymol} shows the density of gas of DM GHOSts (top panel) and classical halos (bottom panel) as a function of radius from the center of mass, with an NFW \citep[e.g.,][]{NFW-96b,NFW-96a,NFW-97} halo profile overplotted (dashed line). 
The stream velocity serves to reduce densities across the structure, as expected. 
Physically, this is due to the advection of gas from the halo and the spatial separation of the two components. 
In addition, at low masses ($\lsim 10^{5.5}$ M$_{\odot}$), we observe a deviation from the NFW shape, with a constant, core profile, rather than a cusp. Using the prolate density profile, Eq.~(\ref{eq:density}), we see that a core-like structure is expected for these ellipsoids (solid line). 
In \S~\ref{sec:nummorphology}, we demonstrated that low mass objects had high eccentricities. 
Since the classical NFW formulation assumes spherical overdensities \citep[e.g.,][]{NFW-96b,NFW-96a,NFW-97}, another reason for the deviation may be the extreme eccentricities of very low mass objects in stream velocity simulations. 

\begin{figure}

 \center

  \includegraphics[width=.48\textwidth]{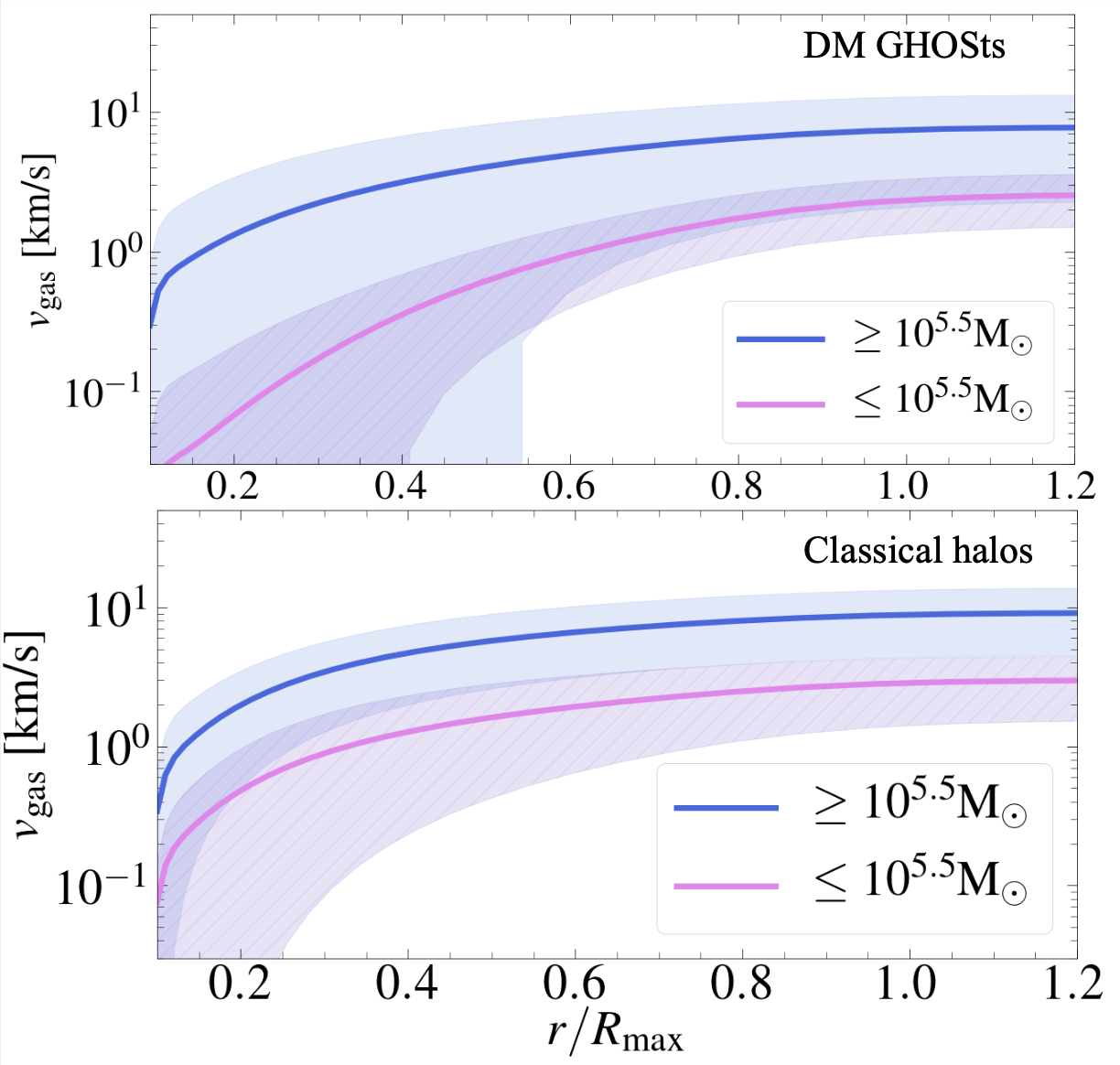}

  \caption{Rotation curves of DM GHOSts as a function of radius, calculated in ellipsoidal shells going outward, for the molecular cooling run with $v_{bc}=2\sigma_{v_{bc}}$ (top panel) and $v_{bc}=0\sigma_{v_{bc}}$ (bottom panel). The average velocity of the gas in an ellipsoidal shell at each radius is normalized by the $v_{circ}$, the circular radius at $R_{\rm max}$ of the ellipsoid. The objects are colored by the mass of their gas component. The average velocity of the gas in an ellipsoidal shell at each radius is normalized by the $v_{circ}$, the circular radius at $R_{\rm max}$ of the ellipsoid.  We split objects by mass--the average rotation curve for objects above $10^{5.5}$ M$_\odot$ is shown in solid blue, while those below this cutoff are plotted in solid pink. 
  Low mass objects display a deviation from the classical rotation curves, with less of a cusp.
  The blue shaded region shows $1\sigma$ away from the curve for objects above $10^{5.5}$ M$_\odot$, while the purple hatched region shows $1\sigma$ away from the curve for objects below $10^{5.5}$ M$_\odot$. 
  }
  \label{fig:rotationmol}

\end{figure}

Figure~\ref{fig:rotationmol} shows the rotation curves of DM GHOSts (top panel) and classical halos (bottom panel). The rotation curves as a function of radius for DM GHOSts have two behaviours separated by mass.
In particular, the core-type density distribution of low-mass objects in the stream also means that their rotational velocity does not climb as fast.

\begin{figure}

 \center

  \includegraphics[width=0.45\textwidth]{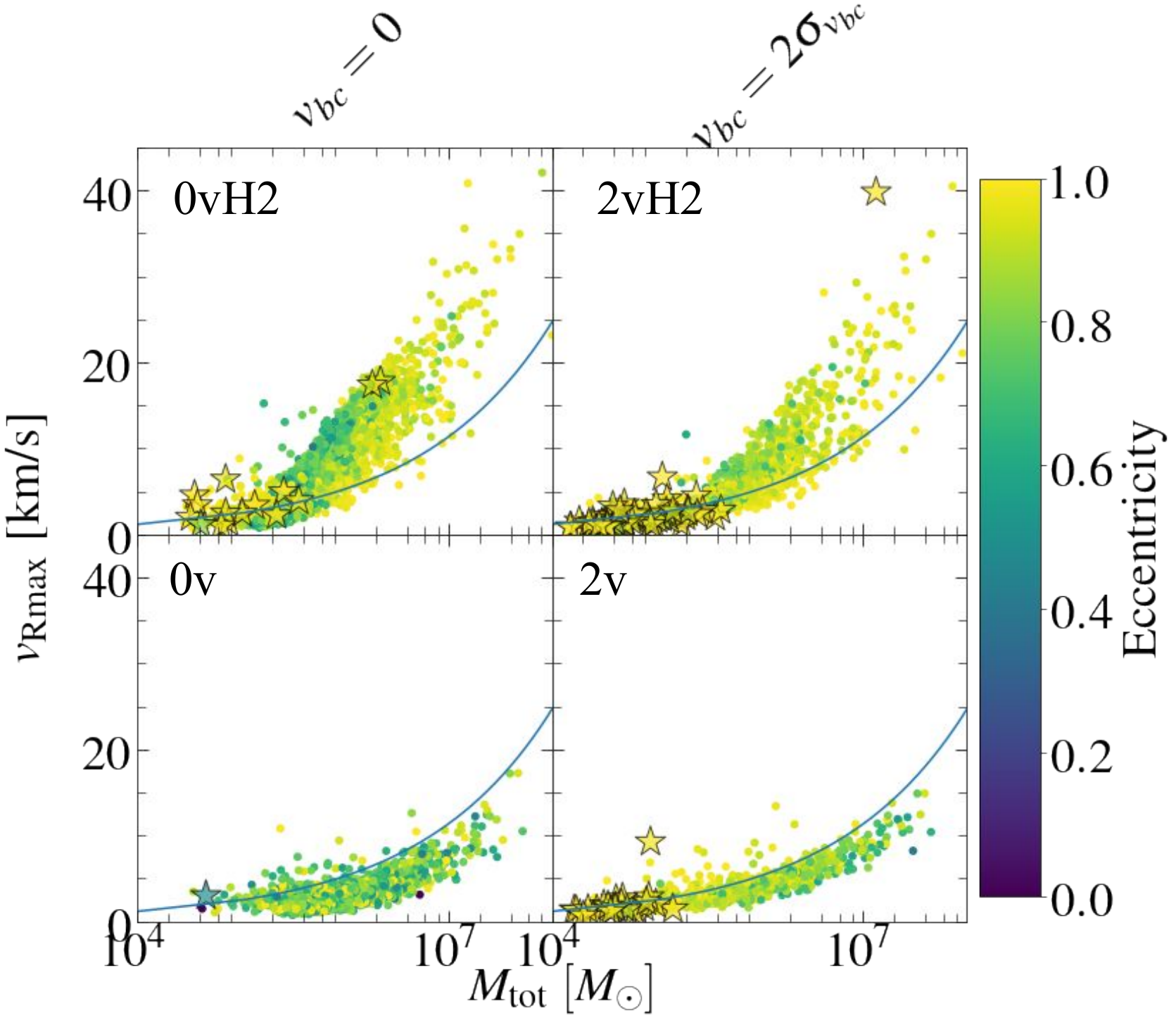}

  \caption{ Scatter plot of the velocity at $R_{\rm max}$ ($v_{Rmax}$) of objects as a function of total mass ($M_{tot}$).  
  The top two panels show the molecular cooling runs, and the bottom two panels show no cooling. 
  The left column has no stream velocity and the right column is from the $v_{bc}=2\sigma_{v_{bc}}$
  Stars represent SIGOs, as defined in Sec.~\ref{sec:objects}. 
  The color bar is the eccentricity (Eq.~(\ref{eq:ecc})).
  The line shows the expected value for an NFW profile.}
  \label{fig:vrmaxmass}

\end{figure}

DM GHOSts are more diffuse and rotationally supported than classical halos, having received a boost from the stream velocity. 
In Figure~\ref{fig:vrmaxmass}, we show the velocity at $R_{\rm max}$ as a function of the total mass. 
We compare this to the nominal NFW expectation, following \citet{NFW-97}, taking the maximum of the NFW circular velocity: 
\begin{equation}
    v_{\rm circ}^2 = \frac{1}{x} \frac{\ln{(1+cx)}-(cx)/(1+cx)}{\ln(1+c)-c/(1+c)},
\end{equation}
where $x=r/r_{200}$, and $c$ is the halo concentration.   
{When molecular cooling is included, the velocity at $R_{\rm max}$ exceeds the NFW circular velocity because the cooling process allows gas to condense to smaller radii. 
In addition to the cooling effect, the stream velocity also boosts the gas velocity.
In regions of streaming, the velocity at $R_{\rm max}$ reaches or exceeds the expected values compared to regions without streaming. 
This behavior is expected based on their larger overall spin parameters than in the classical case as seen in Fig.~\ref{fig:spins}. }

\begin{figure}

 \center

  \includegraphics[width=0.45\textwidth]{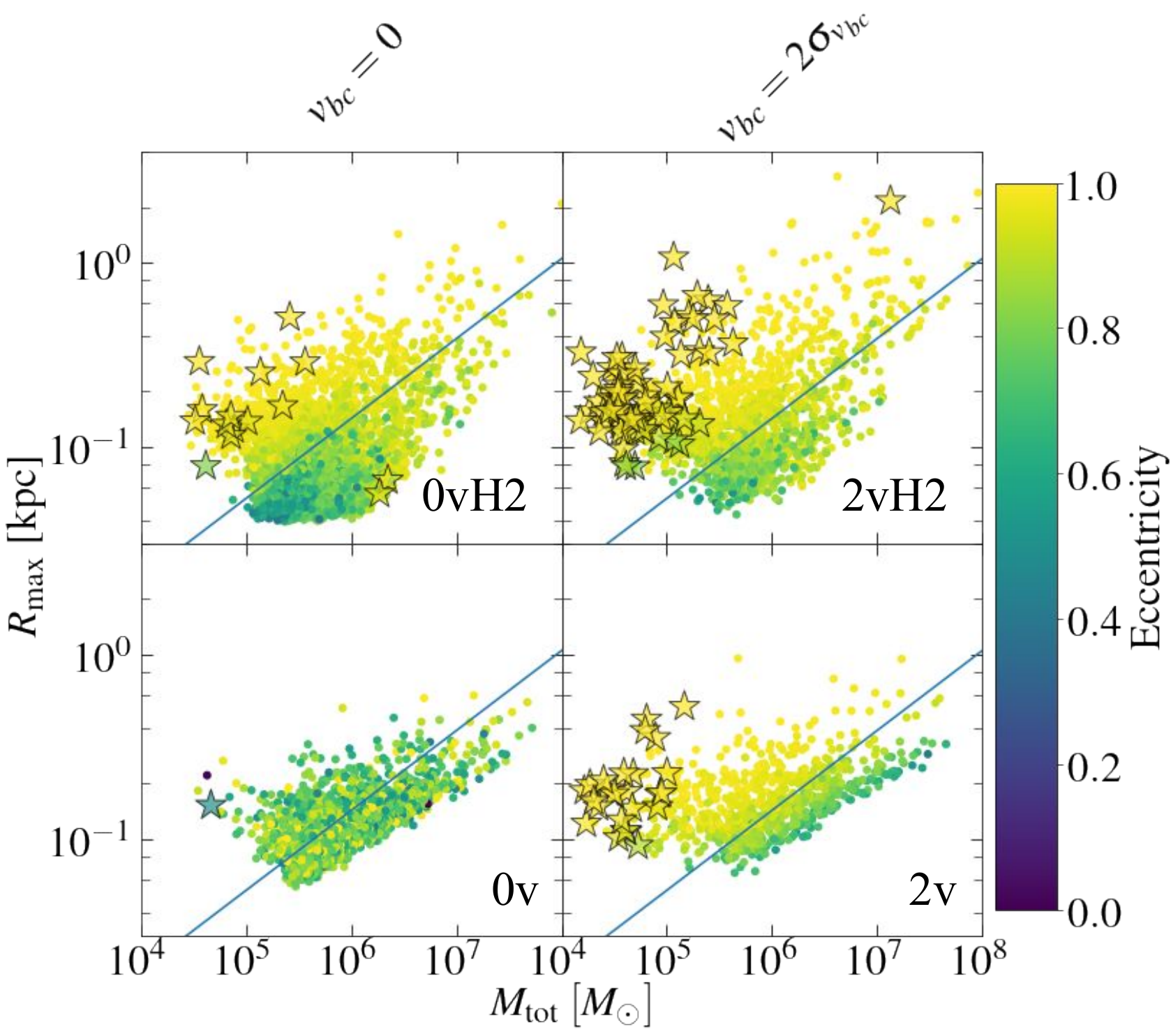}

  \caption{
  Scatter plot of $R_{\rm max}$ of objects as a function of total mass ($M_{tot}$).  
  The top two panels show the molecular cooling runs, and the bottom two panels show no cooling. 
  The left column has no stream velocity and the right column is from the $v_{bc}=2\sigma_{v_{bc}}$
  Stars represent SIGOs, as defined in Sec.~\ref{sec:objects}. 
  There are a few objects misclassified as SIGOs in the runs with no stream velocity--see App.~\ref{ap.gasfraction} for a discussion. 
  The color bar is the eccentricity (Eq.~(\ref{eq:ecc})) 
  The line shows the expected value for an NFW profile.
  For the 0vH2 run, 22\% of objects fall above the expected NFW line. 
  For the 2vH2 run, this fraction rises to 58\% of objects located above the line. 
  }
  \label{fig:rmaxmass}

\end{figure}
We note that the radii of DM GHOSts and SIGOs are larger than the expected from classical considerations. Specifically, in Fig.~\ref{fig:rmaxmass}, the maximum ellipsoid radius is plotted against the total mass of the object, with an NFW expected relationship overplotted.
H$_2$ cooling runs show objects which have condensed to smaller maximum radii (top left panel). 
We calculate the fraction of objects in Fig.~\ref{fig:rmaxmass} above and below the NFW line, and find that the majority ($\sim80\%$) of classical halos condense to smaller radii than the NFW $R_{\rm max}$ with H$_2$ cooling in regions of no stream velocity.   
In the presence of stream velocity, however, the velocity boost overall yields a larger radii (top right panel).
In the 2vH2 run, 60\% of all objects lie $above$ the line, having larger than expected maximum radius. 
Interestingly, SIGOs tend to have higher $R_{\rm max}$ than NFW in all cases. 
Again, we suggest that eccentricity plays a central role in giving objects much greater $R_{\rm max}$ than would be possible in the spherical case. 

These results illustrate the combined effects of the stream velocity and molecular cooling that cause DM GHOSts to be more diffuse and rotationally supported than their classical counterparts of similar masses. 

\section{ Discussion} \label{sec:discussion}

In this work, we investigate the spin, rotational and morphological properties of structures in the presence of stream velocity at $z=20$ using high resolution numerical simulations in {\tt AREPO}. 
For the first time, molecular cooling is included in a detailed study of these dynamical properties. 
We focus on a class of objects that we term DM GHOSts, structures where the baryonic component is  offset from the dark matter halo, but does not fully escape the virial radius (as with SIGOs, which were previously the focus of studies by the Supersonic Project).  
 As in Figure \ref{fig:SIGOdm}, we emphasize that as time goes by, the gas sinks to the center of the DM halo, but carries the signature of its unique formation channel.
Using molecular cooling simulations, we are able to more precisely constrain the properties of SIGOs and DM GHOSts in comparison to classical low mass objects than was possible in previous studies \citep[e.g.,][]{Chiou+18}.

{We considered the following physical properties of DM GHOSts, comparing them to classical objects and SIGOs. }
\begin{itemize}
    \item {\it Morphology:} 
    We show that  SIGOs are the most elongated class of objects, followed by DM GHOSts, {for both gas and DM components} (as depicted in Fig.~\ref{fig:eccentricity} and Fig.~\ref{fig:eccentricityDM}). {We note that the DM component, of DM GHOSts is significantly elongated compared to the classical objects.} 
    Frequently, SIGOs and DM GHOSts tend to be prolate ellipsoidal, and  we present an analytical expression of their gravitational potential. 
    While the gas morphology deviates from spherical symmetry, star formation takes place at density peaks, which end up as less elongated ellipsoids (Lake et al. in prep.). 
    
    Interestingly, we find that the DM component of DM GHOSts is elongated as well, unlike the classical (no stream velocity) counterparts. 
    This prediction may be observable with gravitational lensing models that allow for deviation from spherical symmetry \citep[e.g.,][]{Kneib+11}. 
    {Note that while there is no direct correlation between the stream velocity large scale distribution and the density field, the stream velocity divergence  relates to the density field via the continuity equation \citep[e.g.,][]{Tes+10a,Tes+10b}. Thus, high density $\sigma$ peaks are weakly correlated with large stream velocity patch e.g., \citet[][]{Fialkov14}.  }  
    The box considered here has an increased $\sigma_8$ compared to the average. 
   {Thus, it roughly corresponds to a high redshift progenitor of a patch of the Universe within a density peak such as the Virgo cluster \citep[e.g.,][]{NB07}. Thus, {because of the} above weak correlation, we expect that galaxy clusters are likely to host elongated DM substructures. }
    Thus,  given the right alignments, they may be detected using strong lensing \citep[e.g.,][]{Mahler+22}. {We emphasize that about $40\%$ of the Universe has a stream velocity larger that $1\sigma_{\rm vbc}$, and therefore, DM GOHSts with elongated  gas and DM components should be common regardless of large scale density fluctuations. }

    \item {\it Spin Parameter:} The stream velocity serves to increase the total angular momentum and thus rotational support of SIGOs and DM GHOSts. 
    As shown in Fig.~\ref{fig:spins}, the DM GHOSts have higher gas spin parameter compared to classical objects. 
    Less spherical objects (more eccentric objects) have greater angular momentum, see Fig.~\ref{fig:spinscombined}. 
    
    As expected, the spin vectors of classical gas objects are aligned with those objects' minimum radius, forming a puffy-disk-like configuration at high redshift, consistent with lower redshift analysis for larger objects \citep[e.g.,][]{Jesseit+04,Kautsch+06,Wheeler+17,ElBadry+18}. 
    DM GHOSts, on the other hand, demonstrate spin vectors that are often aligned with their maximum axis (similar to a ``spinning top," see Fig.~\ref{fig:spinsradii}). 
    Lastly,  SIGOs' total gas angular momenta exhibit a weak bifurcation.     
    Most are misaligned with the maximum radius without a preference for alignment with the minimum radius, while another group are aligned towards the maximum radius (similarly to DM GHOSts, as shown in Fig.~\ref{fig:spinsradii}). 
    
    Additionally, the DM and gas components' spins in classical halos are almost always aligned. 
    However,  DM GHOSts, as shown in Fig.~\ref{fig:misalignment}, have a weak preference for alignment between the DM and gas's spin, with a long tail of nearly isotropic configurations. 
    
    \item {\it Mass distribution:} 
    Classical objects are expected to have a cusp-like mass distribution \citep[e.g.,][]{NFW-96b} which are often reproduced in simulations \citep[e.g.,][]{Delos+22}.  
    The stream velocity reduces the density of objects and increases their size, causing them to be puffier and more diffuse than classical objects and the theoretical NFW profile. 
    The ellipsoid-like configuration of low mass DM GHOSts yields a core-like profile (see Fig.~\ref{fig:densitymol}). 
    As expected, SIGOs that follow an ellipsoid profile have a core-like mass density, with a nearly constant density (see Fig~\ref{fig:densitysigos} in App.\ref{ap:morphology}). 
    This behaviour for SIGOs is consistent with the suggestion that SIGOs are giant molecular cloud analogs \citep{Lake+22}. 
    
    \item {\it Rotation curves} The stream velocity affects not only the spin parameter, but also the rotational velocity curves of structures.
    Objects formed by streaming have a higher maximum rotational velocity than those formed without for a given mass (See Figs.~\ref{fig:rotationmol} and~\ref{fig:vrmaxmass}). 
    Furthermore, the bifurcation between high and low mass objects seen in the radial mass distributions for DM GHOSts is also reflected in their velocity profiles. 
    Low mass ($\lsim 10^{5.5}$ M$_{\odot}$) objects, which have cores, do not reach high rotational velocities at their inner radii. 
    The inclusion of molecular cooling increases the velocity at the maximum radius and decreases the maximum radius by condensing rotationally supported material inward. 
 
     We note that rotational curve anomalies have been observed for slightly larger objects in the local Universe \citep[e.g.,][]{Sales+22}. 
     We speculate that anomalous rotation curves produced by the stream velocity at high redshift may persist to low redshift structures. 
     This may be related to the observed ``diversity of rotation curves" problem for ultra faint dwarf galaxies.     
\end{itemize}

The combined effects of molecular cooling and the stream velocity give the most accurate picture to date of the morphological and rotational properties of DM GHOSts. 
We characterize these objects as highly diffuse, rotationally supported dwarf structures with large radii and high eccentricities. 
Based on these anomalous properties, we speculate that at low redshift, DM GHOSts may evolve to form some ultra faint dwarf galaxies or anomalous dwarf galaxies. {In particular, some dwarf galaxies exhibit similar properties, including a diffuse structure and atypical rotation curves \citep[e.g.,][]{BullockBK+17,Sales+22}. 
Thus, while observed ultra faint dwarf galaxies and dwarf galaxies may be more massive than DM GHOSts we find they share similar characteristics at these high redshifts. 
We expect  DM GHOSts to grow over time according to the natural hierarchical growth of structure, and may be the progenitors of some faint dwarf galaxies in regions of the Universe with a highly supersonic stream velocity at early times.}

\section*{Acknowledgements}
The authors would like to thank Sahil Hegde, Bao-Minh Hoang, and Keren Sharon for constructive conversations. 
C.E.W., W.L., S.N., Y.S.C, B.B., F.M., and M.V. thank the support of NASA grant No. 80NSSC20K0500 and the XSEDE AST180056 allocation, as well as the
Simons Foundation Center for Computational Astrophysics and the UCLA cluster \textit{Hoffman2} for computational resources. C.E.W. also thanks the UCLA Competitive Edge program. S.N thanks Howard and Astrid Preston for their generous support. Y.S.C thanks the partial support from UCLA dissertation year fellowship. B.B. also thanks the the Alfred P. Sloan Foundation and the Packard Foundation for support. MV acknowledges support through NASA ATP grants 16-ATP16-0167, 19-ATP19-0019, 19-ATP19-0020, 19-ATP19-0167, and NSF grants AST-1814053, AST-1814259,  AST-1909831 and AST-2007355. NY acknowledges financial support from JST AIP Acceleration Research JP20317829.


\begin{appendix}

\section{Choice of Cutoff Gas Fraction}
\label{ap.gasfraction}
In the first papers by the Supersonic Project that included only adiabatic or atomic cooling \citep[e.g.,][]{Popa+15,Chiou+18,Chiou+19,Chiou+21,Lake+21}, a cutoff gas fraction of $f_g = 0.4$ was chosen for the definition of SIGOs. 
Those studies' statistics for SIGO abundances and properties were thus calculated for objects that were located outside of the virial radius of their parent DM halo and had $f_g=0.4$ within the bounds of the ellipsoid fit described in \S~\ref{sec:objects}.
This choice of gas fraction was a somewhat arbitrary choice, motivated by the fact that it was above the cosmic baryon fraction and close to the stellar fraction of globular clusters \citep[][]{Chiou+18}.
\citet{Nakazato+22} found that in molecular cooling simulations, this choice was too lenient, and resulted in the identification of SIGOs in runs \textit{without} the stream velocity. 
\begin{figure*}

 \center

  \includegraphics[width=\textwidth]{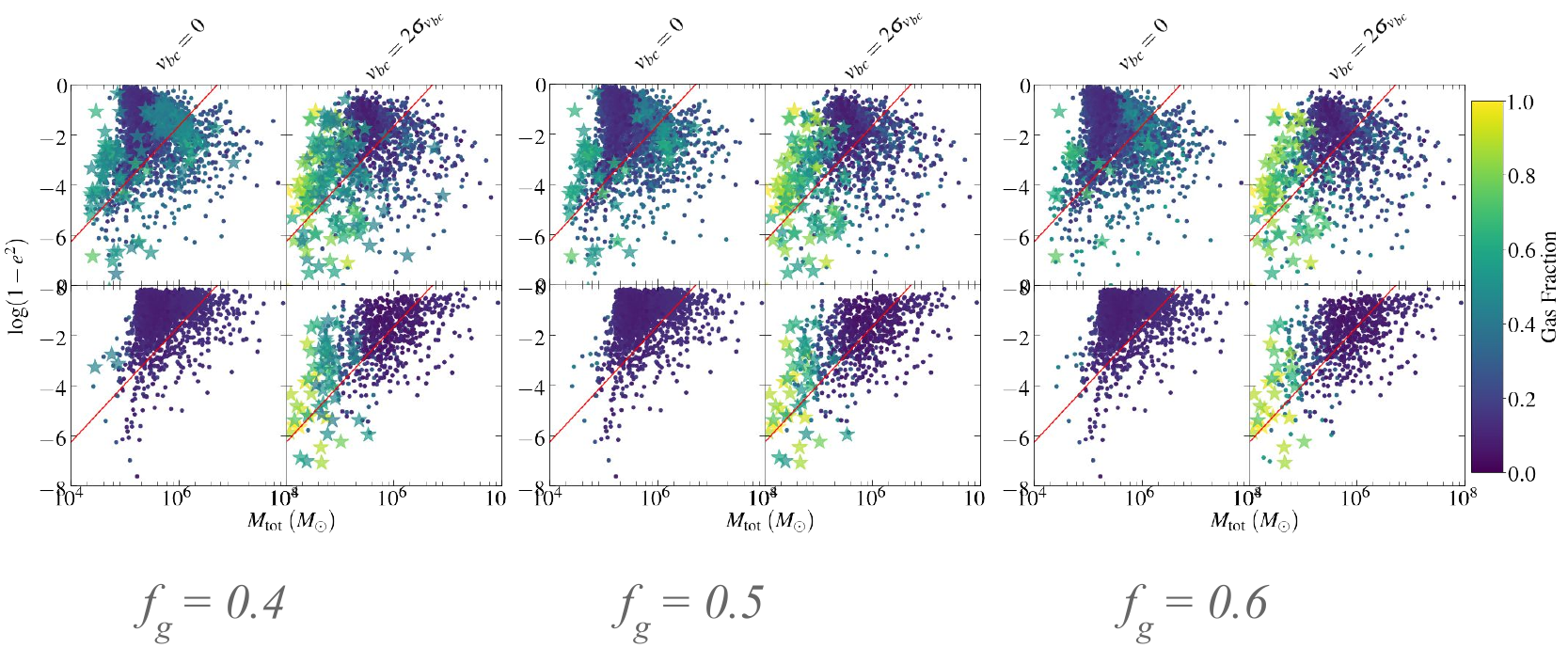}

  \caption{Same as Figure~\ref{fig:eccmass} (Scatter plot of $\log(1-e^2)$ versus $M_{\rm tot}$ for gas-primary (GP) objects,), with the definition of SIGOs calculated using gas fractions of $0.4$ (left), $0.5$ (center) and $0.6$ (right). 
  Significantly more SIGOs are found in the molecular cooling runs without stream velocity (top left of each panel) for $f_g=0.4$ and $f_g =0.5$, as was also shown in \citet{Nakazato+22}. 
  As in Fig.~\ref{fig:eccmass}, the top two panels show the H$_2$ cooling runs, and the bottom two panels show runs without cooling. 
  The left column has no stream velocity and the right column is from the $v_{bc}=2\sigma_{v_{bc}}$ runs. 
  Stars represent SIGOs, as defined in Sec.~\ref{sec:objects}. 
  The color bar is the gas fraction (Eq.~(\ref{eq.gasfraction})).
  The red overplotted line is the expected relationship from Eq.~(\ref{eq:massecc}) for an example object with the average density and maximum radius of objects in the H$_2$ cooling runs ($\bar{\rho}_{Rmax}=1.8\times 10^8 $M$_\odot$ kpc$^{-3}$, $\bar{R}_{max}=0.134$ kpc). }
  \label{fig:gasfraccmasses}

\end{figure*}

We also find that a choice of $f_g=0.4$ results in an unacceptable number of objects being identified as SIGOs in molecular cooling simulations. 
For example, Fig.~\ref{fig:gasfraccmasses} shows the eccentricity versus mass of objects as in Fig.~\ref{fig:eccmass}, with a gas fraction of 0.4 (left), 0.5 (center), and 0.6 (right).  
The top left panel shows the molecular cooling run with no stream velocity, and stars represent SIGOs. 
With $f_g=0.4$ and $f_g=0.5$, there are many objects identified as SIGOs by the algorithm. 
While these gas rich structures may be interesting, they are obviously not the result of a large stream velocity. 
In order to exclude as many of these false SIGOs as possible while still having plenty of objects in the $2$v runs to study, we follow \citet{Nakazato+22} and choose $f_g=0.6$ for this work. 
\begin{figure}

 \center

  \includegraphics[width=0.5\textwidth]{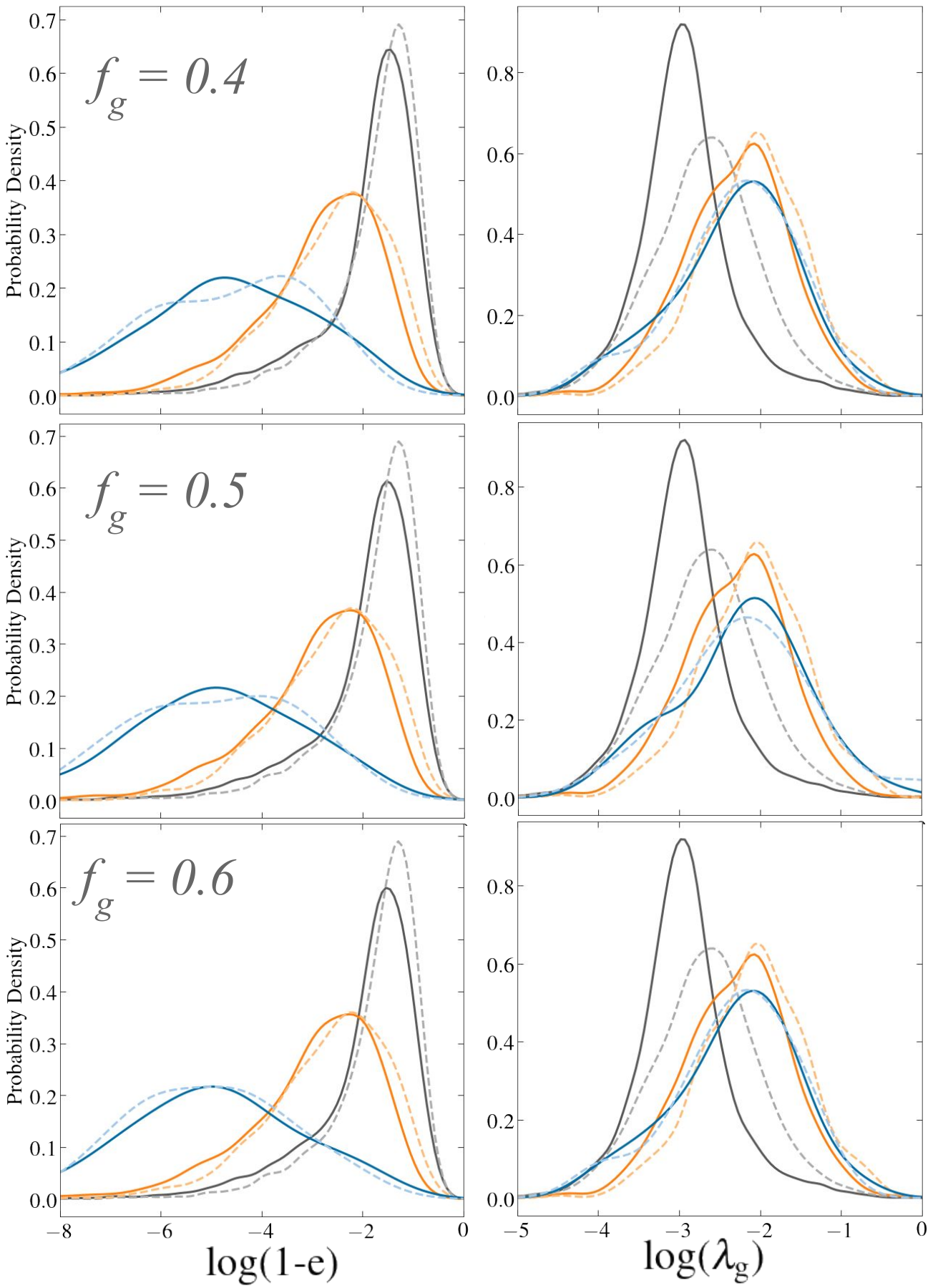}

  \caption{Probability density distributions as shown in Fig.~\ref{fig:eccentricity} ($\log(1-e)$, where $e$ is the eccentricity (Eq.~(\ref{eq:ecc})), for gas primary (GP) objects) and Fig.~\ref{fig:spins} {($\lambda_{\rm g}$ for GP objects),} with the definition of SIGOs calculated using gas fractions of $0.4$ (top row), $0.5$ (center row) and $0.6$ (bottom row). 
  As seen in Fig.~\ref{fig:gasfraccmasses}, more SIGOs are found in the molecular cooling runs without stream velocity for $f_g=0.4$ and $f_g =0.5$, as was also shown in \citet{Nakazato+22}, which motivates us to choose a gas fraction in this work of $0.6$. 
  However, the results are broadly consistent despite variation in gas fraction, and the effects are only seen in the distribution of SIGOs. 
  As in Figs.~\ref{fig:eccmass} and~\ref{fig:spins}, distributions are separated into the object classes listed in Tab.~\ref{Table:objects} and calculated using a Gaussian kernel density. 
 The orange distributions include the gas component of DM GHOSts, the grey distributions show the gas component of classical halos without $v_{bc}$, and the blue distributions show SIGOs. }
  \label{fig:gasfracsparams}

\end{figure}

\begin{figure*}

 \center

  \includegraphics[width=0.75\textwidth]{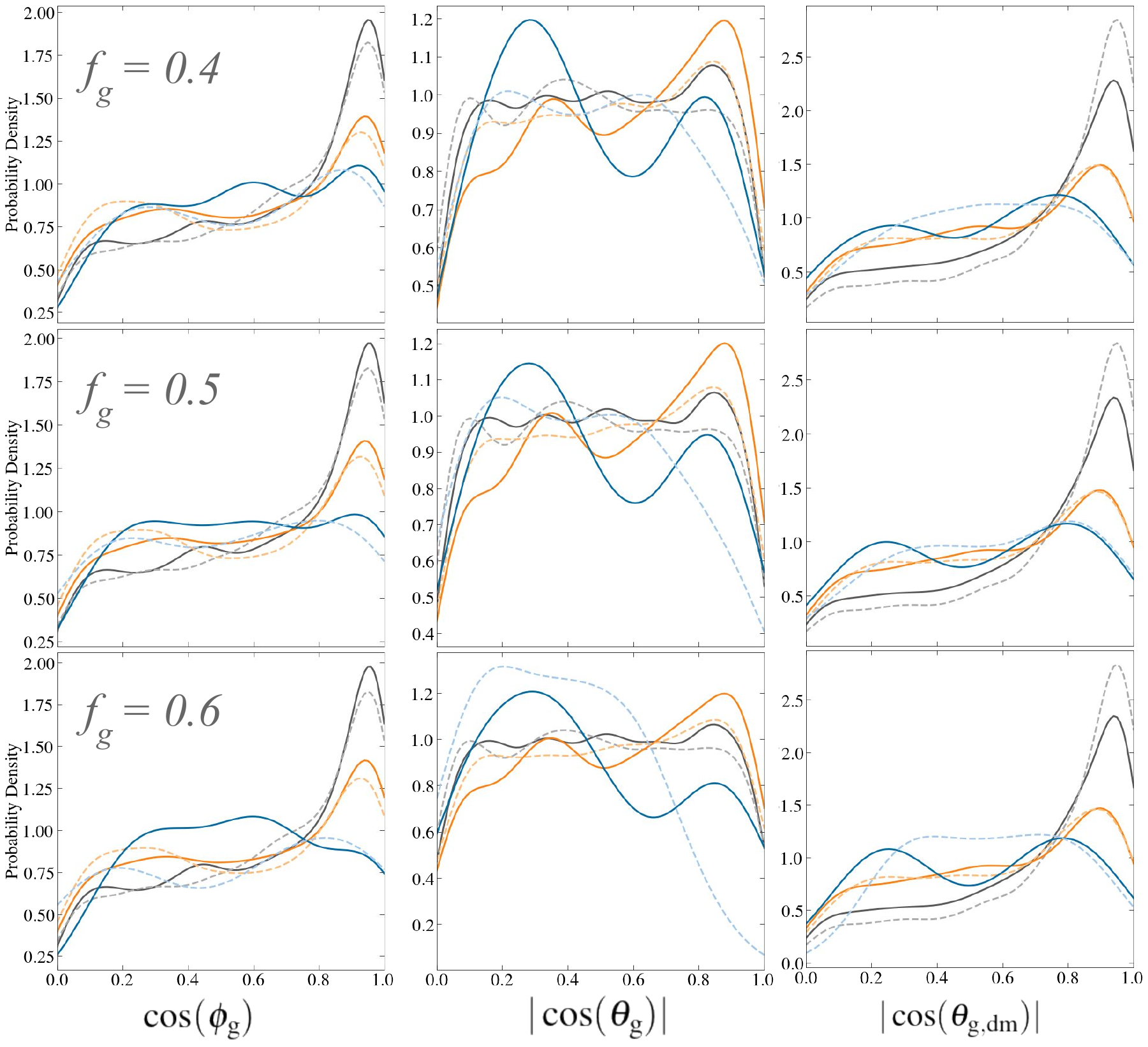}

  \caption{Probability density distributions as shown in Fig.~\ref{fig:spinsradii} (Left: $\cos{\phi_g}$ the angle between rmin and the spin vector of the gas component of  GP objects (See Fig.~\ref{fig:elliosoid}); Center: $|\cos{(\theta_g)}|$, the angle between $R_{\rm max}$ and the spin vector) and Fig.~\ref{fig:misalignment} (Right: $|\cos{\theta_{\rm g,dm}}|$ the misalignment between the spin parameter of the gas and dark matter components of individual objects), with the definition of SIGOs calculated using gas fractions of $0.4$ (top row), $0.5$ (center row) and $0.6$ (bottom row). 
  As seen in Fig.~\ref{fig:gasfraccmasses}, more SIGOs are found in the molecular cooling runs without stream velocity for $f_g=0.4$ and $f_g =0.5$, as was also shown in \citet{Nakazato+22}, which motivates us to choose in this work a gas fraction of $0.6$. 
  However, the results are broadly consistent despite variation in gas fraction, and the effects are only seen in the distribution of SIGOs. 
  As before, the darker lines denote runs with H$_2$ cooling, while the lighter dashed lines denote no cooling.
 Distributions are separated into the object classes listed in Tab.~\ref{Table:objects} and calculated using a Gaussian kernel density. 
 The orange distributions include the gas component of DM GHOSts, the grey distributions show the gas component of classical halos without $v_{bc}$, and the blue distributions show SIGOs.  }
  \label{fig:gasfracsangles}

\end{figure*}

For completeness, Figs.~\ref{fig:gasfracsparams} and~\ref{fig:gasfracsangles} show the probability density distributions for GP objects from this work (as in Figs.~\ref{fig:eccentricity},~\ref{fig:spins},~\ref{fig:spinsradii}, and~\ref{fig:misalignment}) with varying gas fraction from the previous value of $0.4$ to the value of $0.6$ adopted in this work. 
The results are generally consistent despite changing the gas fraction cutoff. 

\section{Derivation of Prolate Ellipsoidal Potential}
\label{ap:potential}
\begin{table}[]
\centering
\begin{tabular}{|l|c|ccc|c|ccc|}
\hline
\textbf{Run} & \textbf{0v} & \multicolumn{3}{c|}{\textbf{2v}} & \textbf{0vH2} & \multicolumn{3}{c|}{\textbf{2vH2}} \\ \hline
\textbf{Objects} & \multicolumn{1}{l|}{\textbf{All}} & \multicolumn{1}{l|}{\textbf{SIGOs}} & \multicolumn{1}{l|}{\textbf{DM GHOSts}} & \multicolumn{1}{l|}{\textbf{All}} & \multicolumn{1}{l|}{\textbf{All}} & \multicolumn{1}{l|}{\textbf{SIGOs}} & \multicolumn{1}{l|}{\textbf{DM GHOSts}} & \multicolumn{1}{l|}{\textbf{All}} \\ \hline
Gas Eccentricity & 7.75E-01 & \multicolumn{1}{c|}{9.82E-01} & \multicolumn{1}{c|}{8.88E-01} & 8.92E-01 & 8.06E-01 & \multicolumn{1}{c|}{9.77E-01} & \multicolumn{1}{c|}{9.11E-01} & 9.15E-01 \\
DM Eccentricity & 8.22E-01 & \multicolumn{1}{c|}{X} & \multicolumn{1}{c|}{8.19E-01} & X & 7.69E-01 & \multicolumn{1}{c|}{X} & \multicolumn{1}{c|}{8.12E-01} & X \\
Gas Spin & 9.12E-02 & \multicolumn{1}{c|}{1.21E-01} & \multicolumn{1}{c|}{1.53E-01} & 1.52E-01 & 6.44E-02 & \multicolumn{1}{c|}{1.80E-01} & \multicolumn{1}{c|}{1.26E-01} & 1.31E-01 \\
DM spin & 3.25E-01 & \multicolumn{1}{c|}{X} & \multicolumn{1}{c|}{2.33E+00} & X & 4.25E-01 & \multicolumn{1}{c|}{X} & \multicolumn{1}{c|}{1.29E+00} & X \\
Total mass (M$_\odot$) & 9.91E+05 & \multicolumn{1}{c|}{3.61E+04} & \multicolumn{1}{c|}{1.75E+06} & 1.69E+06 & 5.39E+05 & \multicolumn{1}{c|}{1.98E+05} & \multicolumn{1}{c|}{1.35E+06} & 1.29E+06 \\
Gas Fraction & 1.20E-01 & \multicolumn{1}{c|}{8.26E-01} & \multicolumn{1}{c|}{1.47E-01} & 1.72E-01 & 2.70E-01 & \multicolumn{1}{c|}{7.72E-01} & \multicolumn{1}{c|}{2.30E-01} & 2.60E-01 \\ \hline
\end{tabular}
\caption{Mean value of selected parameters presented in this work for the four runs. For the $2$v and $2v$H$2$ runs, means are given also for the populations of SIGOs and DM GHOSts separately. }
\label{tab:means}
\end{table}
In this Appendix, we present a derivation of the gravitational potential and total mass of prolate spheroids.
\citet{BT08}  give the general formulae for potentials of various ellipsoidal bodies in their Table 2.1. 
The following equations apply to any inhomogeneous ellipsoid with axes $a_1$, $a_2$ and $a_3$. 
The potential is
\begin{equation}
    \Phi (\mathbf{x}) = -\pi G \frac{a_2 a_3}{a_1}
    \int_0^\infty \frac{\text{d} \tau }{\Delta} \{\varphi(\infty)-\varphi[m(\tau,\mathbf{x})]\},
    \label{eq:potential}
\end{equation}
where
\begin{equation}
    \Delta^2(\tau) \equiv \prod_{i=1}^3 (a_i^2+\tau),
    \label{eq:deltadef}
\end{equation}
\begin{equation}
    m^2(\tau,\mathbf{x})\equiv a_1^2\sum_{i=1}^3\frac{x_i^2}{a_i^2+\tau}
    \label{eq:mdef}
\end{equation}
and
\begin{equation}
    \varphi(m) \equiv \int_0^{m^2}\rho(\mathbf{x})\text{d}m^2(0,\mathbf{x}).
    \label{eq:varphidef}
\end{equation}
For the prolate spheroidal case, we have 
\begin{equation}
    a_1=a_2=R_{\rm min}
\end{equation}
and 
\begin{equation}
    a_3 = R_{\rm max}.
\end{equation}
Thus, in cylindrical coordinates $(R,z)$,
\begin{equation}
    m^2(\tau,\mathbf{x}) = R_{\rm min}^2\left[\frac{R^2}{R_{\rm min}^2+\tau}+\frac{z^2}{R_{\rm max}^2+\tau}\right],
    \label{eq:mproldef}
\end{equation}
and from Eq.~(\ref{eq:deltadef}),
\begin{equation}
    \Delta^2 (\tau)  = (R_{\rm min}^2+\tau)^2(R_{\rm max}^2+\tau).
    \label{eq:deltaproldef}
\end{equation}
Eq.~(\ref{eq:varphidef}) requires a density distribution, and here we will use the following prolate spheroidal density distribution:

\begin{equation}
    \rho(m^2) = \rho_0 \left(1+\left(\frac{m}{a_0}\right)^2\right)^{-\frac{3}{2}},
    \label{eq:appdensity}
\end{equation}
where $\rho_0$ and $a_0$ are constants.

Plugging the density from Eq.~\ref{eq:appdensity} into Eq.~\ref{eq:varphidef} gives: 
\begin{equation}
    \varphi (m) = \int_{0}^{m^2}\rho_0 \left[1+\left(\frac{m}{a_0}\right)^2\right]^{-3/2}\text{d}m^2(0,\mathbf{x})
\end{equation}
\begin{equation}
    =-2a_0^2\rho_0\left[1+\left(\frac{m(0,\mathbf{x})}{a_0}\right)^2\right]^{-1/2}+2a_0^2\rho_0.
    \label{eq:varphim}
\end{equation}
Additionally,
\begin{equation}
    \varphi(\infty) = 2a_0^2\rho_0
    \label{eq:varphiinfty}
\end{equation}
From in Eqs.~(\ref{eq:potential}), ~(\ref{eq:mproldef}),~(\ref{eq:deltaproldef}), ~(\ref{eq:varphim}), and~(\ref{eq:varphiinfty}), the prolate potential is

\begin{equation}
    \Phi (\textbf{x}) = -2\pi GR_{\rm max}a_0^2\rho_0\int_0^{\infty}\frac{\text{d}\tau}{(R_{\rm min}^2+\tau)\sqrt{R_{\rm max}^2+\tau}}\left[1+\left(\frac{m(0,\mathbf{x})}{a_0}\right)^2\right]^{-1/2}
\end{equation}
And with Eq.~(\ref{eq:mproldef}), we have
Equation~(\ref{eq:potential}) thus evaluates to
\begin{equation}
    \Phi (\textbf{x}) = -2\pi GR_{\rm max}a_0^2\rho_0 \int_0^{\infty}\frac{\text{d}\tau}{(R_{\rm min}^2+\tau)\sqrt{R_{\rm max}^2+\tau}}
    \left[1+\left(\frac{R_{\rm min}^2}{a_0^2}\left[\frac{R^2}{R_{\rm min}^2}+\frac{z^2}{R_{\rm max}^2}\right]\right)\right]^{-1/2}
\end{equation}
Evaluating the integral, we get:
\begin{equation}
    \Phi (\textbf{x}) =
    -4\pi GR_{\rm max}^2a_0^3\rho_0  
    \left(\frac{\cos^{-1}\left(\frac{R_{\rm max}}{R_{\rm min}}\right)}{\sqrt{R_{\rm max}^2-R_{\rm min}^2}}\right)
    \frac{1}{\sqrt{1+\left(\frac{R_{\rm min}^2}{a_0^2}\left[\frac{R^2}{R_{\rm min}^2}+\frac{z^2}{R_{\rm max}^2}\right]\right)}}
    \label{eq:finalpotential}
\end{equation}
Taking $a_0=R_{\rm max}$:
\begin{equation}
 \Phi (\textbf{x}) =
    -\frac{4\pi GR_{\rm max}^4 \rho_0\cos^{-1}(\sqrt{1-e^2})}{e\sqrt{1+(1-e^2)\left(\frac{R^2}{R_{\rm min}^2}+\frac{z^2}{R_{\rm max}^2}\right)}}.
    \label{eq:finalpotentiala0}
\end{equation}

Now, we find the dependence of the total mass on eccentricity, starting with a similar argument to that presented in \citet{BT08} for the potential of oblate spheroids. 
In cylindrical coordinates a prolate spheroidal shell with axes $\beta R_{\rm max}$ and $\beta R_{\rm min}$ is given by:
\begin{equation}
    \frac{R^2}{R_{\rm min}^2}+\frac{z^2}{R_{\rm max}^2}=\beta^2.
\end{equation}
Here, the $z$-axis is aligned with the polar radius ($R_{\rm max}$) and the $R$ coordinate points in the direction of the equatorial radius ($R_{\rm min}$).  
The volume enclosed inside this shell is given by
\begin{equation}
    V = \frac{4}{3}\pi R_{\rm max}R_{\rm min}^2 \beta^3
\end{equation}
\begin{equation}
    = \frac{4}{3}\pi R_{\rm max}^3 \beta^3 (1-e^2)
\end{equation}
Thus, assuming a constant surface density, the mass enclosed between two shells $\beta$ and $\beta+\delta\beta$ is:
\begin{equation}
    \delta M = 4\pi \rho R_{\rm max}^3 (1-e^2)\beta^2 \delta \beta
    \label{eq:massdifferential}
\end{equation}
The full mass of the ellipsoid is found by integrating over a set of similar spheroids from the center to the outer edge of the object. 
Using the notation of \citet{BT08}, this set is given by all the spheroids for which: \begin{equation}
    \text{constant} = m^2 \equiv \frac{R^2}{1-e^2}+z^2.
\end{equation}
This constant $m=\beta R_{\rm max}$. 
Thus, for some density function $\rho(m^2)$, according to Eq.~(\ref{eq:massdifferential}), 
\begin{equation}
    \delta M = 4\pi \rho(m^2)(1-e^2)m^2 \delta m.
\end{equation}
Integrating this equation over the ellipsoid gives the total mass: 
\begin{equation}
   M = 4 \pi (1-e^2) \int_0^{R_{\rm max}} \rho(m^2)m^2 dm.
   \label{eq:generalintegral}
\end{equation}
Once again, we assume the density distribution of Eq.~(\ref{eq:density}), and plugging into Eq.~(\ref{eq:generalintegral}), we solve 
\begin{equation}
    M = 4 \pi (1-e^2) \int_0^{R_{\rm max}} \left(1+\left(\frac{m}{a_0}\right)^2\right)^{-\frac{3}{2}} m^2 dm.
\end{equation}
to obtain
\begin{equation}
    M= 4 \pi \rho_0 (1-e^2) a_0^3
     \left[\sinh^{-1}{\left(\frac{R_{\rm max}}{a_0}\right)}-\frac{R_{\rm max}}{\sqrt{a_0^2+R_{\rm max}^2}}\right].
\end{equation}
Taking $a_0 = R_{\rm max}$ as above gives:
\begin{equation}
    M=4\pi \rho_0(1-e^2) R_{\rm max}^3 \left(\sinh^{-1}{(1)}-\frac{1}{\sqrt{2}}\right)
    \label{eq:masseccap}
\end{equation}
\begin{equation*}
    \approx 2.19 \rho_0 (1-e^2)R_{\rm max}^3.
\end{equation*}

\section{Morphological Investigation}
\label{ap:morphology}
In this Appendix, we include several supporting Figures relating to our morphological and rotational investigation above. 
Table~\ref{tab:means} lists the means of selected distributions from this work. 

\begin{figure}

 \center

  \includegraphics[width=0.8\textwidth]{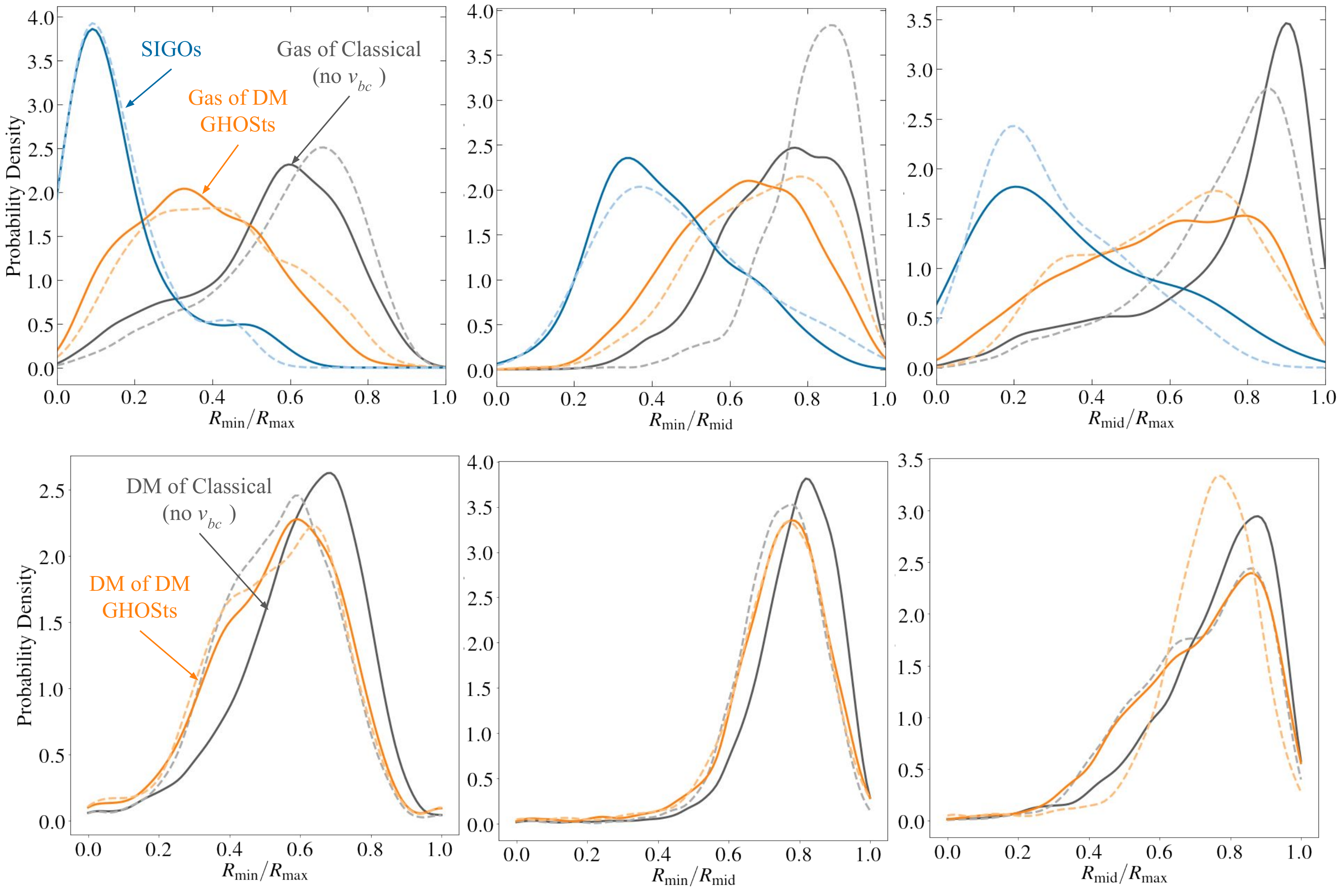}

  \caption{Probability density distribution of the ratios of GP objects. 
Following the same convention as in Fig.~\ref{fig:eccentricity}, the various distributions demonstrate runs with and without the stream velocity and cooling. 
 The orange distributions show objects with a gas fraction $f$ of less than 0.6 in the runs without a stream velocity ($v_{bc}=0\sigma_{v_{bc}}$), the grey distributions show the classical equivalent of $f<0.6$ objects in the $v_{bc}=0\sigma_{v_{bc}}$ run, and the blue dashed distributions show SIGOs, which are only found in the $v_{bc}=2\sigma_{v_{bc}}$ run. 
 The darker lines denote no cooling, while the lighter dashed lines show the inclusion of molecular cooling. }
  \label{fig:ratios}

\end{figure}
\begin{figure}

 \center

  \includegraphics[width=0.5\textwidth]{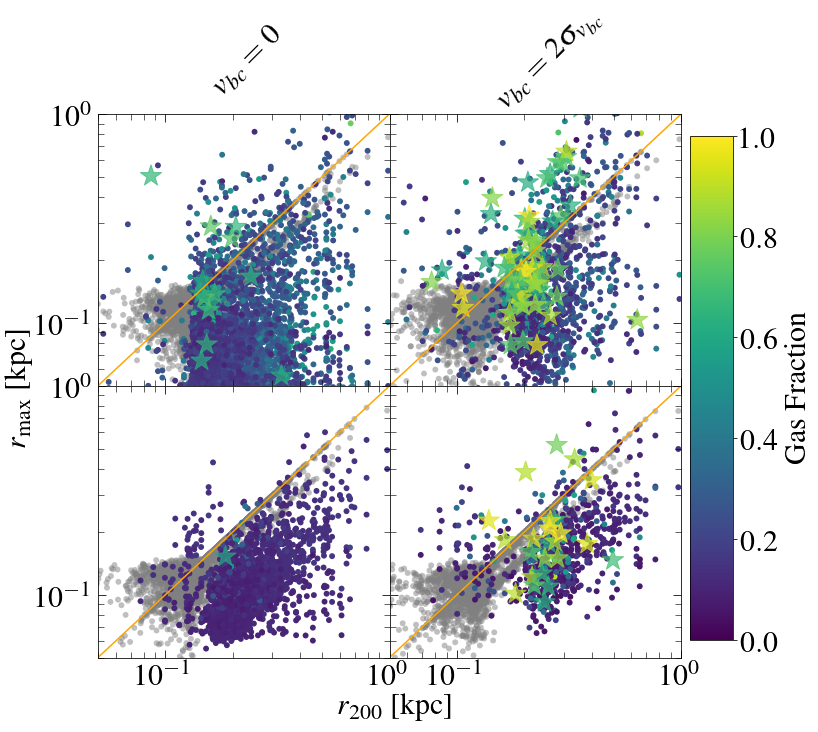}

  \caption{Scatter plot of $R_{\rm max}$ ($v_{Rmax}$) versus $R_{200}$ of GP objects. 
  The top two panels show the molecular cooling runs, and the bottom two panels show no cooling. 
  The left column has no stream velocity and the right column is from the $v_{bc}=2\sigma_{v_{bc}}$.
  Stars represent SIGOs, as defined in Sec.~\ref{sec:objects}. 
  The color bar is the gas fraction (Eq.~(\ref{eq.gasfraction})) 
  The orange line shows $R_{\rm max}=R_{200}$. Colored points are GP objects, and grey points are DM/G objects.  }
  \label{fig:rmaxr200}
 \end{figure}
In Fig~\ref{fig:ratios}, we plot the probability density distributions of the three axis ratios plotted on the axes of Fig.~\ref{fig:eccentricity}.
The classical halos tendency towards sphericity 
($R_{\rm min}/R_{\rm mid}\sim R_{\rm min}/R_{\rm max}\sim R_{\rm mid}/R_{\rm max}\sim 1$) is clearly seen here, as well as the distinct deviation of SIGOs and DM GHOSts away from sphericity. 
For SIGOs especially, the distributions show evidence of triaxiality (($R_{\rm min}/R_{\rm mid}\neq R_{\rm min}/R_{\rm max}\neq R_{\rm mid}/R_{\rm max}$). 
Figure~\ref{fig:ratios} also shows the axes ratios for DM primary objects (bottom row). 
Here, we see that the DM components of DM GHOSts are not only more eccentric, having a tail of small axes ratios as in Fig.~\ref{fig:eccentricity}, but also show prolate shapes when $R_{\rm mid}$ is taken into account. 
In the center bottom panel, the ratio $R_{\rm min}/R_{\rm mid}$ for the DM component is close to one, whereas the ratio $R_{\rm mid}/R_{\rm max}$ has a tail of small values. 
This is an indication of prolateness ($R_{\rm min}\sim R_{\rm mid}<R_{\rm max}$).

  \begin{figure}

 \center

  \includegraphics[width=0.5\textwidth]{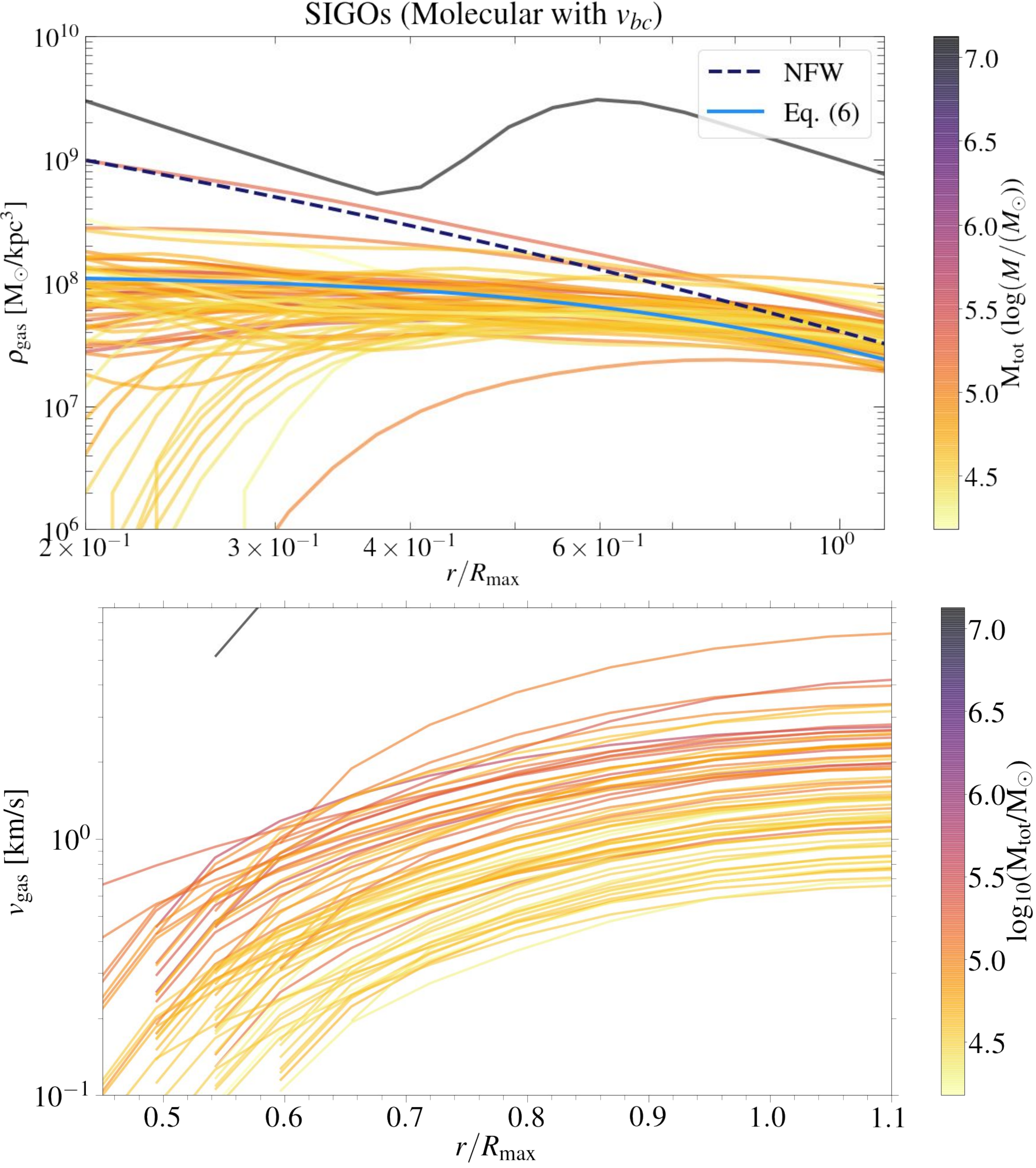}

  \caption{\textbf{Top panel}: Density of gas in SIGOs as a function of radius, normalized to the $R_{\rm max}$ of the ellipsoid, for the molecular cooling run with $v_{bc}=2\sigma_{v_{bc}}$.
  The density is calculated in 50 ellipsoidal shells moving out from the center of the object. The objects are colored by the mass of their gas component.
  An NFW profile for a $10^5$ M$_\odot$ halo is shown for comparison as the dashed line. 
  Eq.~(\ref{eq:density}) is plotted for a SIGO with average density in solid blue.
  \textbf{Bottom panel:} Rotation curves of SIGOs as a function of radius, calculated in ellipsoidal shells going outward, for the molecular cooling run with $v_{bc}=2\sigma_{v_{bc}}$.
  The average velocity of the gas in an ellipsoidal shell at each radius is normalized by the $v_{circ}$, the circular radius at $R_{\rm max}$ of the ellipsoid. 
  The objects are colored by the mass of their gas component. 
  The average velocity of the gas in an ellipsoidal shell at each radius is normalized by the $v_{circ}$, the circular radius at $R_{\rm max}$ of the ellipsoid. 
  The objects are colored by the mass of their gas component. 
  The average SIGO has $\bar{\rho}_{\rm Rmax}=4.14\times 10^{7}$ M$_\odot$ kpc$^{-3}$ and $\bar{R}_{\rm max}=0.240 $ kpc}
  \label{fig:densitysigos}
\end{figure}
In Fig.~\ref{fig:rmaxr200}, we plot the maximum gas ellipsoid radius against the $R_{200}$ of its parent halo. 
The DM halos, in general, are much larger than the gas--this is expected. 
The stream velocity (as mentioned in \S~\ref{sec:nummorphology}) drives more extreme eccentricity, leading to large $R_{\rm max}$.

The DM maximum ellipsoid radius is also shown as dark points in Fig.~\ref{fig:rmaxr200} as a function of the $R_{200}$ found from a spherical overdensity calculation. 
The orange line corresponds to $R_{\rm max,DM}=R_{200}$
In Fig.~\ref{fig:eccmass}, it was shown that low mass objects have higher eccentricity. 
This is reflected in the fact that the DM distribution deviated from the $1:1$ line at low masses, whereas most DM objects fall on the line at higher masses.

Finally, in Fig.~\ref{fig:densitysigos}, we show the radial gas density and rotation curve of all the SIGOs from the 0vH2 run, with an NFW profile and Eq.~(\ref{eq:density}) overplotted. 
As with the low mass DM GHOSts, the NFW is not a good fit. 
SIGOs seem to have a core, rather than a cusp.

\end{appendix}


\newpage
\bibliography{cosmo}{}
\bibliographystyle{aasjournal}
\end{document}